\documentclass[aps,prl,reprint,superscriptaddress,letterpaper,twocolumn,longbibliography]{revtex4-2}
\usepackage[english]{babel}
\usepackage{ucs}
\usepackage[utf8x]{inputenc}
\usepackage{amsmath}
\usepackage{mathtools}
\usepackage{amssymb}
\usepackage{amsthm}
\usepackage{mathrsfs}
\usepackage{graphicx}
\usepackage{bm}
\usepackage{bbm}
\usepackage{empheq}
\usepackage{cases}
\usepackage{euscript}
\usepackage[usenames, dvipsnames, x11names]{xcolor}
\usepackage[colorlinks=true,linkcolor=SpringGreen4,citecolor=blue,urlcolor=Magenta3]{hyperref}
\usepackage[shortlabels]{enumitem}
\usepackage[todonotes={textsize=scriptsize}]{changes}

\definechangesauthor[name={Hugo Ribeiro}, color=SpringGreen4]{HR}
\definechangesauthor[name={Florian Marquardt}, color=Blue4]{FM}
\newcommand{\norm}[1]{\left\lVert#1\right\rVert}
\newcommand{\mm}[1]{\mathrm{#1}}
\newcommand{\abs}[1]{\left|#1\right|}
\newcommand{\drv}[2]{\frac{\ud#1}{\ud#2}}

\newcommand{\di}[1]{\mathop{}\!\mathrm{d} #1}

\newcommand{\s}[1]{{\sigma}_{#1}}

\def \uc{\mathrm{c}}
\def \ud{\mathrm{d}}

\def \uf{\mathrm{f}}

\def \um{\mathrm{m}}

\def \uG{\mathrm{G}}
\def \uI{\mathrm{I}}
\def \uL{\mathrm{L}}
\def \uM{\mathrm{M}}

\def \cv{\mbox{\boldmath$c$}}

\def \rv{\mbox{\boldmath$r$}}

\def \ev{\mbox{\boldmath$e$}}

\def \ev{\mbox{\boldmath$e$}}
\def \vv{\mbox{\boldmath$v$}}
\def \xv{\mbox{\boldmath$x$}}
\def \yv{\mbox{\boldmath$y$}}
\DeclareFontFamily{OT1}{pzc}{}
\DeclareFontShape{OT1}{pzc}{m}{it}{<-> s * [1.10] pzcmi7t}{}
\DeclareMathAlphabet{\mathpzc}{OT1}{pzc}{m}{it}

\begin{document}

\title{Accelerated Non-Reciprocal Transfer of Energy Around an Exceptional Point}

\author{Hugo Ribeiro}
\affiliation{Max Planck Institute for the Science of Light, Staudtstraße 2, 91058 Erlangen, Germany}

\author{Florian Marquardt}
\affiliation{Max Planck Institute for the Science of Light, Staudtstraße 2, 91058 Erlangen, Germany}
\affiliation{Institute for Theoretical Physics, Department of Physics, University of Erlangen-Nürnberg, Staudtstrasse 7, 91058
Erlangen, Germany}

\begin{abstract}
	We develop perturbative methods to study and control dynamical phenomena related to exceptional points in Non-Hermitian
	systems. In particular, we show how to find perturbative solutions based on the Magnus expansion that accurately describe
	the evolution of non-Hermitian systems when encircling an exceptional point. This allows us to use the recently proposed
	Magnus-based strategy for control to design fast non-reciprocal, topological operations whose fidelity error is
	orders-of-magnitude smaller than their much slower adiabatic counterparts.
\end{abstract}

\maketitle

\textit{Introduction ---} 
A peculiar feature of non-Hermitian systems~\cite{el-ganainy2018} is the existence in parameter space of branch point
singularities at which two or more eigenvalues, and their corresponding eigenvectors, coalesce and become
degenerate~\cite{kato1995,berry2004,seyranian2005,heiss2012}. The existence of such singular points, known as exceptional points,
in the spectrum of a non-Hermitian system has led in recent years to the development of novel functionalities in optics and
photonics systems~\cite{lin2011,regensburger2012,feng2013,feng2014,hodaei2014,peng2014,weimann2017,miri2019,ozdemir2019} and to
reconsider the understanding of topological quantum matter~\cite{xiao2020,bergholtz2021}.

In particular, it was predicted~\cite{heiss1999,heiss2000,keck2003,berry2011,uzdin2011,milburn2015} and demonstrated in
several platforms~\cite{dembowski2001,xu2016,doppler2016,liu2021} that enclosing an exceptional point via a slow
varying closed loop results in a non-reciprocal exchange of energy between the two normal modes of the system. The exchange of
energy is non-reciprocal both  with respect to the initial condition and orientation of the control loop.

The operations generated by enclosing an exceptional point are topological; the control loop defines a closed path
in parameter space that cannot be continuously deformed to a single point without crossing the singularity. However, as
one speeds up the rate at which the parameters defining the control loop vary, the topological properties vanish. This is
analogous to chiral edge transport in periodic photonic structures, where transport is robust against disorder and
imperfections, but only if the amount of disorder and imperfections is weak enough~\cite{li2009,gonzalez2019}.

Here, we present a perturbative method based on the Magnus expansion~\cite{magnus1954,blanes2009} that allows one to describe the
evolution of non-Hermitian systems. As we show below, the perturbative solutions accurately predict the dynamics when an
exceptional point is encircled by a closed control loop. Furthermore, the existence of perturbative solutions allows us to design
closed control loops that are both fast and more effective at exchanging the energy between the modes than their slower
counterparts while simultaneously keeping the topological, non-reciprocal character of the operation. To achieve
this goal, we build on the recently proposed Magnus-based strategy for control~\cite{ribeiro2017,roque2021} that we
extend to the problem of non-reciprocal dynamics.

\begin{figure}[t!]
        \includegraphics[width=\columnwidth]{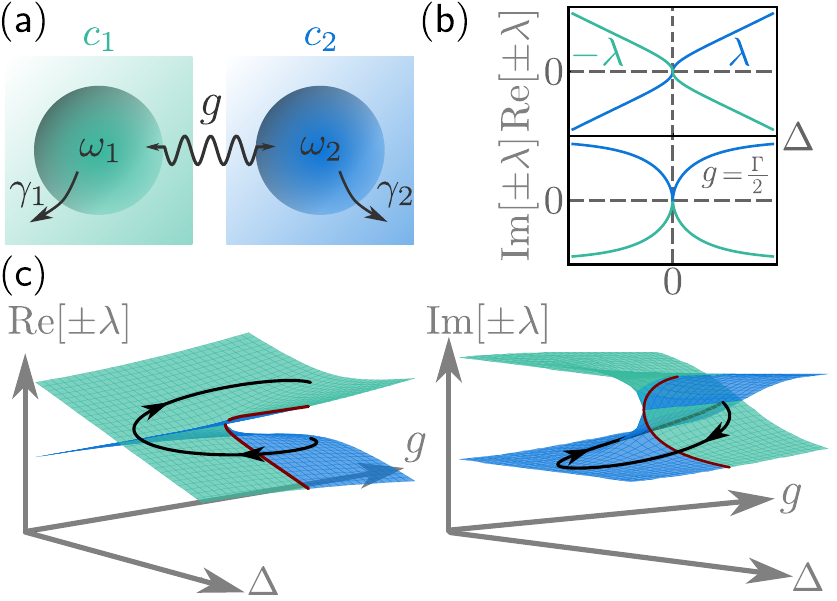}
	\caption{
		Exceptional points in a dissipative coupled two-mode system. (a) Schematic representation of two coupled modes
		with dissipation. (b) Cross sections of the eigenvalue surfaces for $g = \Gamma/2$ as a function of detuning
		$\Delta$, showing the location of an exceptional point located at $(\Delta =0, g = \Gamma/2)$. (c) Real
		and imaginary parts of the eigenvalues in the two-dimensional parameter space $(\Delta, g)$ with a negatively
		oriented closed path in parameter space about the exceptional point. The red cross section emphasizes when the
		path crosses from the manifold defined by one eigenvalue to the other.
		}
	\label{fig:fig01} 
\end{figure}

\textit{Dynamics around an exceptional point ---} 
We consider two coupled harmonics modes with time-dependent frequencies $\omega_1 (t)$ and $\omega_2 (t)$, coupling strength $g
(t)$, and without loss of generality time-independent decay rates $\gamma_1$ and $\gamma_2$ [see Fig.~\ref{fig:fig01}~(a)]. The
dynamical system that describes the evolution of the mode amplitudes is given by
\begin{equation}
	\dot{\Phi}_\mm{sym} (t) = -i D_\mm{sym} (t) \Phi_\mm{sym} (t),
	\label{eq:LabFrameDynSys}
\end{equation}
where $D_\mm{sym} (t) = -(\Delta (t) + i \Gamma/2) \s{z,\mm{sym}} + g (t) \s{x,\mm{sym}}$ and $\Phi_\mm{sym} (t)$ is the flow
from which we can find the modes amplitude vector $\cv_\mm{sym} (t) = [c_{\mm{sym},1} (t), c_{\mm{sym},2}, (t)]^\mathsf{T}$ at
time $t$, i.e, $\cv_\mm{sym} (t) = \Phi_\mm{sym} (t) \cv_\mm{sym} (0)$, and which obeys the initial condition $\Phi_\mm{sym} (0) =
\mathbbm{1}$. We have defined $\Delta (t) = [\omega_1 (t) - \omega_2 (t)]/2$, $\Gamma = (\gamma_1 - \gamma_2)/2$, and
$\s{j,\mm{sym}}$ with $j\in \{x,y,z\}$ are Pauli matrices.

It is convenient to work in the frame that diagonalizes $D_\mm{sym} (t)$ at each instant in time (adiabatic frame). This is done
via the change-of-frame transformation $S(t) = \exp(-i \theta (t) \s{y})$ with $\theta (t) = \arctan[-g(t) / (\Delta (t) + i
\Gamma/2)]/2$, i.e., $\Phi_\mm{sym} (t) \to \Phi (t)$ $=S^{-1} (t) \Phi_\mm{sym} (t) S (0)$, where we assume that the
evolution starts at $t=0$. The flow $\Phi (t)$ describes the evolution of the normal modes and obeys the equation of motion 
\begin{equation}
	\dot{\Phi} (t) = -i D(t) \Phi (t) = -i \left(\lambda (t) \s{z,\mm{ad}} - \dot{\theta} (t) \s{y,\mm{ad}}\right)  \Phi (t).
	\label{eq:EigFrameDynSys}
\end{equation}
Since the change-of-frame matrix $S(t)$ is explicitly time-dependent, transforming Eq.~\eqref{eq:LabFrameDynSys} to the adiabatic
frame generates a non-inertial coupling term (non-adiabatic coupling) between the normal modes $\cv_+ = (1,0)^\mathsf{T}$ and
$\cv_- = (0,1)^\mathsf{T}$ with strength $\dot{\theta} (t)$. The instantaneous, complex eigenvalues of $D_\mm{sym} (t)$ associated
to the eigenmodes $\cv_+$ and $\cv_-$ are $\pm \lambda (t)$, respectively, with $\lambda (t) = \sqrt{(\Delta (t) + i \Gamma/2)^2 +
g^2 (t)}$. 

Exceptional points in the spectrum of $D_\mm{sym} (t)$ occur at $(\Delta =0, g = \pm \Gamma/2)$, where the two eigenvalues
coalesce ($\lambda (t) =0$) [see Fig.~\ref{fig:fig01}~(b)].

We are interested in the dynamics described by Eq.~\eqref{eq:EigFrameDynSys} when one of the exceptional points is enclosed by a
control loop. We consider closed control loops of duration $t_\uf$, enclosing the exceptional point located at $(\Delta =0, g =
\Gamma/2)$. An example of such a control loop is the circular path parametrized by
\begin{equation}
	\rv (t) = \left[ r_0 \sin\left( \frac{2 \pi s}{t_\uf} + \alpha \right),\,\frac{\Gamma}{2} + r_0
	\cos\left(\frac{2 \pi s}{t_\uf} t + \alpha\right)\right],
	\label{eq:CircPathParam}
\end{equation}
where $r_0$ is the radius of the circle, $s = +1 (\circlearrowright)$, $-1 (\circlearrowleft)$ defines the orientation, and
$\alpha$ parametrizes the starting point of the loop. 

\begin{figure}[t!]
        \includegraphics[width=\columnwidth]{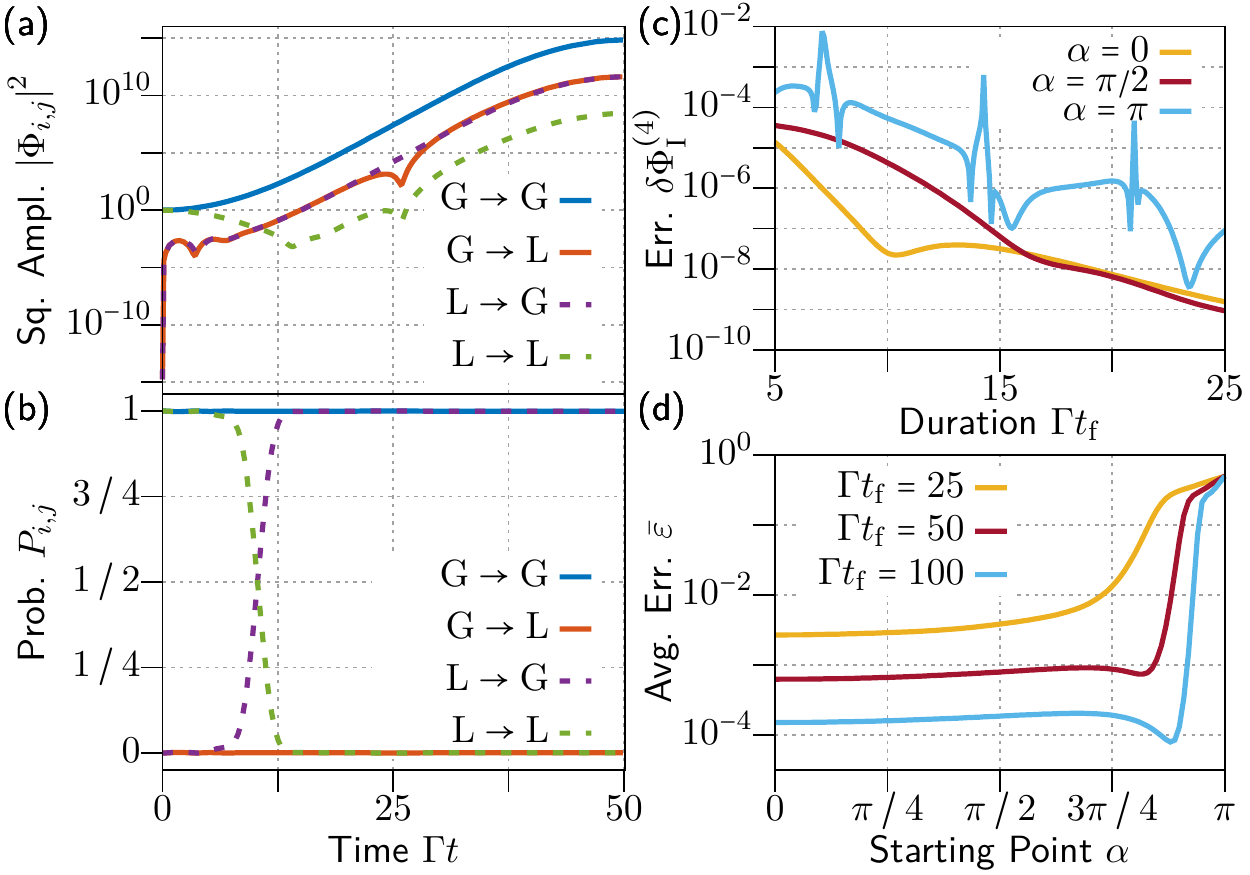}
	\caption{
		Non-reciprocal dynamics with (uncorrected) closed control loops [see Eq.~\eqref{eq:CircPathParam}]
		and fidelity error of the Magnus approximation. (a) Evolution of the squared matrix elements of the flow $\Phi
		(t)$ [see Eq.~\eqref{eq:EigFrameDynSys}] and (b) their normalized counterparts [see Eq.~\eqref{eq:Prob}]. (c)
		Time-averaged error of the approximate solution [see Eq.~\eqref{eq:ApproxSolErr}]. (d) Average error of the
		non-reciprocal exchange of energy [see Eq.~\eqref{eq:AvgErr}] as function of the starting point of the control
		loop.  Unless otherwise specified, we chose  $\Gamma t_\uf =50$, $r_0 =
		(1/2)\Gamma$, and $\alpha=0$. 
		}
	\label{fig:fig02} 
\end{figure}

In what follows we show how to get approximate solutions of Eq.~\eqref{eq:EigFrameDynSys} using the Magnus
expansion~\cite{magnus1954,blanes2009}. Using perturbation theory is particularly challenging in this context due
to the amplification dynamics generated by Eq.~\eqref{eq:EigFrameDynSys}, which exponentially amplifies small
perturbations. While one might think that this will inevitably lead to perturbation theory to break down, we show that
this is not the case, if the perturbative expansion is done in a suitable frame.

We look for solutions of Eq.~\eqref{eq:EigFrameDynSys} of the form 
\begin{equation}
	\Phi (t) = \Phi_0(t) \Phi_\uI (t),
	\label{eq:IntFramePhi}
\end{equation}
where 
\begin{equation}
	\Phi_0 (t) =  \exp\left(-i \Gamma t_\uf \left\{\mm{Re}[\tilde{\Lambda} (t)] + i \mm{Im}[\tilde{\Lambda}
	(t)]\right\} \sigma_{z,\mm{ad}}\right)
	\label{eq:Phi0}
\end{equation}
is a solution of Eq.~\eqref{eq:EigFrameDynSys} where the non-adiabatic coupling, i.e., the term proportional to $\dot{\theta}
(t)$, is fully neglected. We have defined 
\begin{equation}
	\Lambda (t) = \int_0^t \di{t_1} \lambda (t_1) 
	= t_\uf \int_0^{\frac{t}{t_\uf}} \di{x} \lambda (x) 
	= \Gamma t_\uf \tilde{\Lambda}(t)
	\label{eq:EigInt}
\end{equation}
with the second equality following from the change of variable $x=t/t_\uf$.  Within this framework, the flow $\Phi_\uI (t)$ can
then be interpreted as the deviation from the ideal adiabatic dynamics described by $\Phi_0 (t)$. 

The flow $ \Phi_0 (t)$ predicts that the amplitude of one of the eigenmodes is amplified while the amplitude of the other mode is
damped with the sign of $\mm{Im}[\Lambda (t)]$ determining which of the eigenmodes undergoes instantaneous amplification and
damping, respectively. Therefore, it is convenient to introduce the gain mode $\cv_\uG =\delta_{1,f(t_\uf,s)} \cv_+ +
\delta_{-1,f(t_\uf,s)} \cv_-$ and lossy mode $\cv_\uL = (1-\delta_{1,f(t_\uf,s)}) \cv_+ + (1 - \delta_{-1,f(t_\uf,s)}) \cv_-$,
where  $\delta_{i,j}$ denotes the Kronecker delta function and we have defined $f(t_\uf, s) = \mm{sign} \{\mm{Im}[\Lambda
(t_\uf)]\}$. The gain mode (lossy mode) is associated to the eigenmode whose amplitude is amplified (damped) at $t=t_\uf$
according to the prediction of $\Phi_0 (t)$.

Since the flow $\Phi_0 (t)$ is diagonal, it cannot describe the non-reciprocal dynamics, which is our main concern here. As
previously identified (see, e.g., Ref.~\cite{milburn2015}), the non-adiabatic coupling is a necessary ingredient to generate
non-reciprocal dynamics. This is best understood when considering the dynamical equation for $\Phi_\uI (t)$, which is obtained by
substituting Eq.~\eqref{eq:IntFramePhi} into Eq.~\eqref{eq:EigFrameDynSys}. We have
\begin{equation}
	\dot{\Phi}_\uI (t) =
	D_\uI (t) \Phi_\uI (t) =  
	\dot{\theta} (t) \left(e^{2 i \Lambda (t)} \sigma_+ - e^{-2 i \Lambda (t)} \sigma_- \right) \Phi_\uI (t),
	\label{eq:EigFrameDynSysI}
\end{equation}
where we have introduced the matrices $\sigma_\pm = (\s{x} \pm i \s{y})/2$. 

We recall that for a \emph{Hermitian system}, i.e., $\Lambda (t) \in \mathbb{R}$, we would have $D_\uI (t) \to \mathbf{0}$ as
$t_\uf \to \infty$ since $\dot{\theta} (t) \propto 1/t_\uf$. This would yield $\Phi_\uI (t) \to \mathbbm{1}$ and there would be no
deviations from the ideal adiabatic dynamics. 

In stark contrast to the Hermitian case, $D_\uI (t) \not \to \mathbf{0}$ as $t_\uf \to \infty$ for \emph{all} times since at least
one matrix element of $D_\uI (t)$ can be exponentially large in $t_\uf$. This readily follows from substituting
Eq.~\eqref{eq:EigInt} into Eq.~\eqref{eq:EigFrameDynSysI} and splitting $\tilde{\Lambda} (t)$ into real and imaginary parts. As a
consequence, even in the long-cycling limit, there are always deviations from the ideal adiabatic dynamics, which illustrates that
the adiabatic theorem~\cite{born1928} does not hold for non-Hermitian systems in general. 

We approximate the solutions of Eq.~\eqref{eq:EigFrameDynSysI} using a Dyson series~\cite{dyson1949}. More specifically, we use
the relation between the Magnus expansion~\cite{magnus1954,blanes2009} and the Dyson series (see Section 2.4 in
Ref.~\cite{blanes2009}) to represent the solutions as 
\begin{equation}
	\Phi_\uI (t) =  \exp\left[ \sum_{k=1}^\infty \epsilon^k \Omega_k (t) \right] 
	=  \mathbbm{1} + \sum_{j=1}^\infty \frac{1}{j!} \left[ \sum_{k=1}^\infty \epsilon^k \Omega_k (t) \right]^j,
	\label{eq:PhiIApprox}
\end{equation}
where $\Omega_k (t)$ is the $k$th term of the Magnus series (see, e.g, Ref.~\cite{blanes2009}) and we have introduced the
parameter $\epsilon$ for bookkeeping. In the following, we denote by $\Phi_\uI^{(n)} (t)$ the truncated series expansion where we
keep at most terms of order $n$, i.e, $\Phi_\uI (t) = \Phi_\uI^{(n)} (t) + \mathcal{O}(\epsilon^{n+1})$. 

To assess the quality of the approximation, we define the time-averaged error 
\begin{equation}
	\delta \Phi_\uI^{(n)} = \abs{1 - \frac{1}{6} \sum_{j \in S} \frac{1}{t_\uf} \int_0^{t_\uf} \di{t}
	\abs{\left[ \ev_j^{(n)} (t) \right]^\mathsf{H} \cdot \ev_j (t)}^2}
	\label{eq:ApproxSolErr}
\end{equation}
where $\vv^\mathsf{H}$ denotes the conjugate transpose of the vector $\vv$ and  we have introduced the unit vectors $\ev_i (t) =
\Phi (t) \cv_i (0)/\norm{\Phi (t) \cv_i (0)}$ and $\ev_i^{(n)} (t) =  \Phi^{(n)} (t) \cv_i (0)/\norm{\Phi^{(n)} (t) \cv_i (0)}$.
The quantity $F_i (t) = \abs{[\ev_i^{(n)} (t)]^\mathsf{H} \cdot \ev_i (t)}^2$ gives the state dependent fidelity at time $t$ between
the approximated unit state vector $\ev_i^{(n)} (t)$ and the exact unit state vector $\ev_i (t)$. We get the time-averaged
fidelity by averaging over time and over the six initial states $\cv_i (0)$ with $i\in S=\{\pm x, \pm y, \pm z\}$. These six
initial states correspond to the eigenvectors of the Pauli matrices.

In Fig.~\ref{fig:fig02}~(c), we plot $\delta \Phi_\uI^{(4)}$ for the control loop defined in Eq.~\eqref{eq:CircPathParam} as a
function of the duration $\Gamma t_\uf$. The results show that our perturbative solutions of Eq.~\eqref{eq:EigFrameDynSys} based
on the Magnus expansion accurately describe the dynamics of the system. We have made the perturbative expansion possible by
choosing an appropriate interaction picture, where the size of the perturbation remains relatively small compared to the generator
of the unperturbed dynamics.

The approximate solutions provide an intuitive way to understand how the interplay between non-adiabatic transitions and
amplification leads to the non-reciprocal behavior with respect to the initial condition. In the limit $\Gamma t_\uf \gg 1$, we
find that the matrix elements of the flow $\Phi (t)$ behave asymptotically, i.e., for $\Gamma t \gg 1$, according to (see
Supplemental Material)
\begin{equation}
	\begin{aligned}
		\abs{\Phi^{(2)}_{\uG,\uG} (t)}^2 &\sim e^{2 \abs{\mm{Im}[\Lambda (t)]}} 
		+ \mathcal{O}\left[ \left( \Gamma t_\uf \right)^{-1} \right], \\
		\abs{\Phi^{(2)}_{\uL,\uG} (t)}^2 &\sim e^{2 \abs{\mm{Im}[\Lambda (t)]}}
		\abs{\dot{\theta} (t) / [2 \lambda (t)]}^2 + \mathcal{O}\left[ \left( \Gamma t_\uf \right)^{-3}\right], \\
		\abs{\Phi^{(2)}_{\uG,\uL} (t)}^2 &\sim e^{2 \abs{\mm{Im}[\Lambda (t)]}} 
		\abs{\dot{\theta} (0) / [2 \lambda (0)]}^2 + \mathcal{O}\left[ \left( \Gamma t_\uf \right)^{-3}\right], \\
		\abs{\Phi^{(2)}_{\uL,\uL} (t)}^2 &\sim e^{2 \abs{\mm{Im}[\Lambda (t)]}}
		\abs{\dot{\theta} (0) \dot{\theta} (t) / [4 \lambda (0) \lambda (t)]}^2 
		+ \mathcal{O}\left[ \left( \Gamma t_\uf \right)^{-5}\right],
	\end{aligned}
	\label{eq:ApproxSolEls}
\end{equation}
where $\Phi_{i,j} (t) = \cv_i^\mathsf{T} \Phi (t) \cv_j$ with $i,\,j\in{\uG,\uL}$ and we have assumed $\abs{\mm{Im}[\Lambda
(t_\uf)]} \gg 1$. The result shows that the net effect of non-adiabatic transitions can be reduced to what happens
at the boundaries of the time-interval over which the evolution takes place. This is in complete analogy to the Hermitian
case~\cite{wiebe2012} and can similarly be derived using the Magnus expansion~\cite{ribeiro2019}.

Furthermore, Eq.~\eqref{eq:ApproxSolEls} shows that all transmission channels undergo amplitude amplification asymptotically [see
Fig.~\ref{fig:fig02}~(a)]. While this might seem counter-intuitive at first, especially for the $\uL \to \uL$ channel, it
directly follows from an interplay between the amplitude of the gain mode being amplified and non-adiabatic transitions. For
instance, if the system is initialized in the lossy mode, i.e., $\cv (0) = \cv_L$, amplitude is going to be transferred via the
non-adiabatic coupling to the gain mode at $t=0$, where it will be amplified, only to return at a later time $t$ back to the lossy
mode.

Using Eq.~\eqref{eq:ApproxSolEls}, we can evaluate asymptotically the ratio $\eta (t)$ between the energy stored in the lossy mode
and the gain mode. We find 
\begin{equation}
	\eta (t) = \frac{\abs{\Phi_{\uL,\uG} (t)}^2}{\abs{\Phi_{\uG,\uG} (t)}^2} 
	= \frac{\abs{\Phi_{\uL,\uL}(t)}^2}{\abs{\Phi_{\uG,\uL} (t)}^2} 
	\sim \frac{1}{4}\frac{\abs{\dot{\theta}(t)}^2}{\abs{\lambda (t)}^2} \propto \frac{1}{4}\frac{1}{(\Gamma t_\uf)^2},
	\label{eq:Efficiency}
\end{equation}
which tends to $0$ as $t_\uf \to \infty$. This result, which is independent of the initial state, indicates that most of the
energy ends up in the gain mode, with the latter being determined by the orientation of the control loop. This is the expected
non-reciprocal behavior, which is best observed when considering the normalized squared amplitudes
\begin{equation}
	P_{i,j} (t) = \frac{\abs{\Phi_{i,j} (t)}^2}{\sum_{i=\uG}^\uL \abs{\Phi_{i,j} (t)}^2},
	\label{eq:Prob}
\end{equation}
as shown in Fig.~\ref{fig:fig02}~(b).

In summary, there are two necessary conditions to fulfill to realize a highly efficient non-reciprocal energy transfer:~(i) the
dynamics must generate a sizeable amount of amplification for all transmission channels $i \to j$ $(i,\,j\in\{\uG, \uL\})$, which
is identical to requiring $\abs{\mm{Im}[\Lambda (t_\uf)]} \gg 1$, and (ii) the ratio between the energy stored in the lossy mode
and the gain mode should be small at $t=t_\uf$, i.e., $\eta (t_\uf) \ll 1$ [see Eq.~\eqref{eq:Efficiency}]. The latter condition
ensures a highly efficient transfer since most of the energy ends up in the gain mode at the end of the control loop.

While the  efficiency is not contingent on the choice of a specific closed contour, it depends on the duration of
the control loop [see Eq.~\eqref{eq:Efficiency}] and on the starting point of the control loop. The latter can be
understood geometrically by noticing that a closed contour in parameter space does not correspond to a closed contour on
the Riemannian manifold defined by the real and imaginary parts of the spectrum [see Fig.~\ref{fig:fig01}~(c)]. Thus,
changing the starting point of the control loop in parameter space can lead to paths on the Riemannian manifold of the
spectrum that result in $\mm{Im}[\lambda (t)]$ being an anti-symmetric function of time around $t=t_\uf/2$. Such a
situation leads to $\mm{Im}[\Lambda (t_\uf)] = 0$ for which condition (i) does not hold. Thus, enclosing an exceptional
point with a slow varying control loop does not always lead to non-reciprocal dynamics.

We illustrate this behavior in Fig.~\ref{fig:fig02}~(d) by plotting for fixed $\Gamma t_\uf$ the average error
\begin{equation}
	\bar{\varepsilon} = 1 - \frac{1}{4}\sum_{j=\pm} \left[P_{+,j}^\circlearrowright (t_\uf) +
	P_{-,j}^\circlearrowleft (t_\uf) \right],
	\label{eq:AvgErr}
\end{equation}
calculated for the control loop defined in Eq.~\eqref{eq:CircPathParam} as a function of $\alpha$. We recall that $\alpha$
parametrizes the position of the starting point for the circular loop in parameter space. For $\alpha=\pi$, the error becomes
maximal because the non-reciprocity is broken due to having $\mm{Im}[\Lambda (t_\uf)] = 0$.

We have defined the average error such that $\bar{\varepsilon}=0$ corresponds to a perfect non-reciprocal transfer of energy,
i.e., all of the energy is transferred to the gain mode. We have expressed the normalized squared amplitudes in
Eq.~\eqref{eq:AvgErr} in the basis of eigenmodes and explicitly indicated the path orientation.

\textit{The Control Problem ---}
We are now in a position to show how to design control loops that lead to a highly efficient non-reciprocal exchange of energy
even when the cycling time becomes small, i.e., $\Gamma t_\uf \sim 1$. Our approach follows from the recently proposed
Magnus-based strategy for control introduced in Refs.~\cite{ribeiro2017,roque2021}.

\begin{figure}[t!]
	\includegraphics[width=\columnwidth]{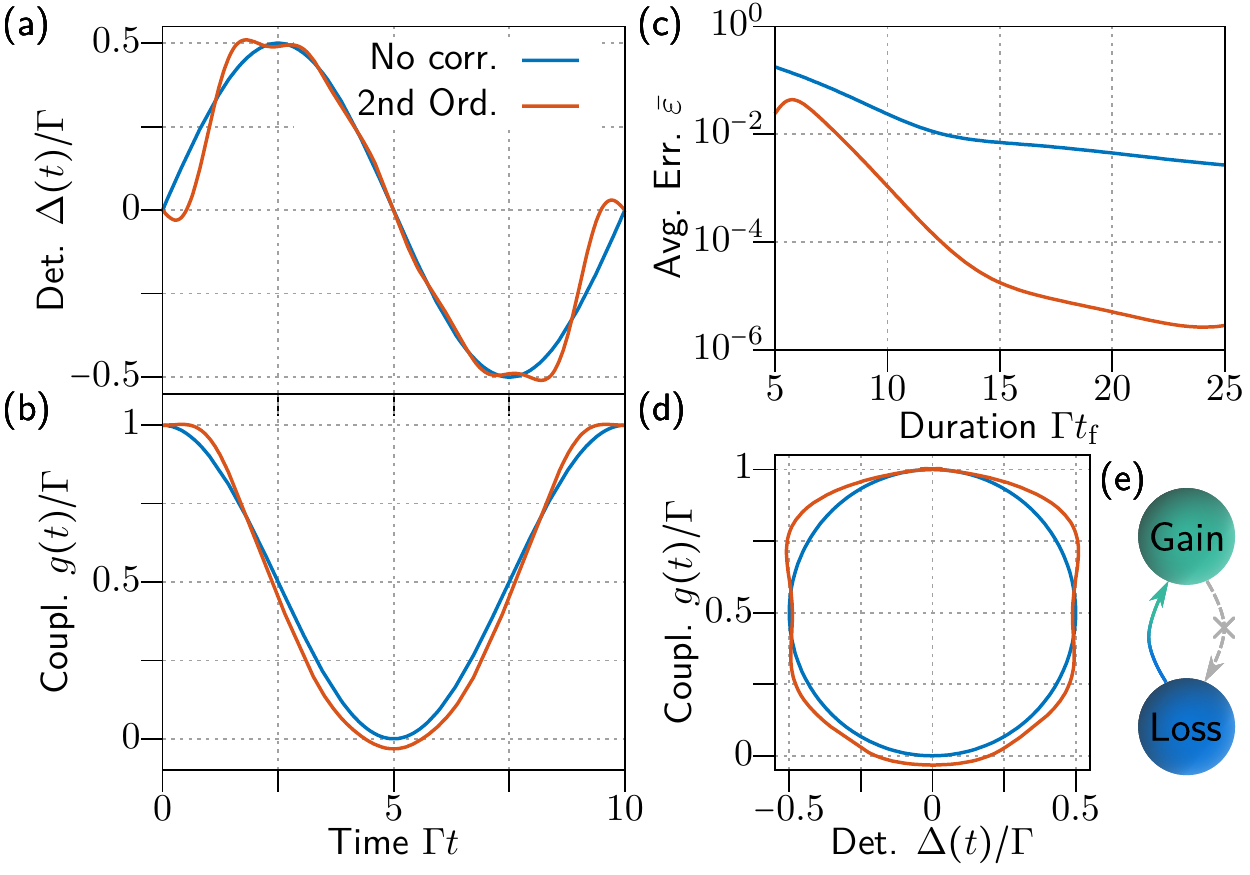}
	\caption{
		Accelerated non-reciprocal exchange of energy via corrected control loops around an exceptional
		point. (a) - (b) Comparison between initially chosen and second-order modified control fields [see
		Eq.~\eqref{eq:DeltaAndgCorr}] and (d) resulting path in parameter space. (c) Average error of the non-reciprocal
		exchange of energy [see Eq.~\eqref{eq:AvgErr}], suppressed by orders of magnitude. (e) Schematic representation of
		the \emph{average} non-adiabatic transitions induced by the modified dynamics. Unless specified, we chose $\Gamma
		t_\uf=10$, $r_0 = (1/2)\Gamma$, $\alpha =0$, $k_\um = k_\uM$ $=$ $n_\um = n_\uM=0$, $l_\um = m_\um =1$, and $l_\uM
		= m_\uM =6$.
		}
	\label{fig:fig03} 
\end{figure}

The first step of the Magnus-based strategy for control entails finding a partition of the dynamical matrix $D(t) =  D_\mm{ideal}
(t) + V_\mm{bad} (t)$, where $D_\mm{ideal} (t)$ generates a flow $\Phi_\mm{ideal} (t) = T\exp[-i \int_0^t \di{t_1} D_\mm{ideal}
(t_1)]$ such that $\Phi_\mm{ideal} (t_\uf) = \Phi_\uG$ is the desired operation one wishes to realize at $t=t_\uf$.  $V_\mm{bad}
(t)$ is the spurious dynamical matrix disrupting the ideal dynamics and preventing one to achieve the desired operation. 

The second step consists in modifying $D (t)$ by introducing a control $W (t)$ which on average cancels the deleterious effects
generated by $V_\mm{bad} (t)$, i.e., $D (t) \to D_\mm{mod} (t) = D(t) + W(t)$. Formally, $W (t)$ describes how the time dependence
of $D(t)$ needs to be modified to generate the desired operation at $t=t_\uf$.

For the problem at hand, we have $D_\mm{ideal} (t) = \lambda (t) \sigma_z + s i \dot{\theta} (t) (\sigma_x + s i \sigma_y)/2$,
which depends explicitly on the orientation $s$ of any chosen path. Finding an exact closed form representation for
$\Phi_\mm{ideal} (t)$ is a challenging task, thus forcing one to rely on numerical approaches. While this is possible, it negates
one of the main advantages of the Manugs-based strategy for control, which is the ability to treat the control problem
semi-analytically.

In the following, we propose a modified way to partition the original control problem, and which can also be employed in the
quantum case, when an exact representation for $\Phi_\mm{ideal} (t)$ is hard to find. In contrast to the prescription of
Refs.~\cite{ribeiro2017,roque2021}, we decompose $D_\mm{mod} (t)$ into
\begin{equation}
	D_\mm{mod} (t) = D_\mm{easy} (t) + V_\mm{good} (t) + V_\mm{bad} (t) + W(t),
	\label{eq:DMod}
\end{equation}
where $D_\mm{easy} (t) + V_\mm{good} (t) = D_\mm{ideal} (t)$. The decomposition introduced in Eq.~\eqref{eq:DMod} is chosen such
that it is straightforward to find the flow $\Phi_\mm{easy} (t)$ generated by $D_\mm{easy} (t)$. 

To find $W (t)$, it is convenient to transform Eq.~\eqref{eq:DMod} to the interaction picture defined by $\Phi_\mm{easy} (t)$ and
represent $W (t)$ as a series, i.e., $W(t) = \sum_n W^{(n)} (t)$. We obtain 
\begin{equation}
	D_\mm{mod,I} (t) =  V_\mm{good,I} (t) + V_\mm{bad,I} (t) + \sum_n W^{(n)}_\uI (t).
	\label{eq:DModI}
\end{equation}

Following the philosophy of Refs.~\cite{ribeiro2017,roque2021}, we can determine $W^{(n)}_\uI (t)$ by considering the Magnus
expansion generated by the partially corrected dynamical matrix $D_\mm{mod,I}^{(n)} (t) = V_\mm{good,I} (t) + V_\mm{bad,I} (t) +
\sum_{l=1}^n W_\uI^{(l)} (t)$. Taking into account that $W^{(n)}_\uI (t)$ must only cancel on average spurious terms involving
$V_\mm{bad,I} (t)$, we find that $W^{(n)}_\uI (t)$ must satisfy the following equations:
\begin{equation}
	\int_0^{t_\uf} \di{t} W^{(n)}_\uI (t) = -i \sum_{l=1}^n \left[ \Omega_\mm{bad,l}^{(n-1)} (t_\uf) + \delta \Omega_l^{(n-1)}
	(t_\uf)\right],
	\label{eq:EqWI}
\end{equation}
with $\Omega_l^{(n)} (t) = \Omega_{\mm{good},l}^{(n)} (t) + \Omega_{\mm{bad},l}^{(n)} (t) + \delta \Omega_l^{(n)} (t)$ the $l$th
term of the Magnus expansion associated to $D_\mm{mod,I}^{(n)} (t)$
\footnote{For $n=0$, we define $D_\mm{mod,I}^{(0)} (t) = V_\mm{good,I} (t) + V_\mm{bad,I} (t)$.}.
We have decomposed $\Omega_l^{(n)} (t)$ into contributions coming from $V_\mm{good,I} (t)$, $V_\mm{bad,I} (t)$, and commutators
involving both $V_\mm{good,I} (t)$ and $V_\mm{bad,I} (t)$, respectively. 

Using the decomposition introduced in Eq.~\eqref{eq:DMod}, we do not need to find explicitly $\Phi_\mm{ideal} (t)$, which can be
hard, and deal with the difficult, good part and bad part of the dynamics using the Magnus expansion instead. This allows us to
use the Magnus-based strategy for control as it was intended to be used:~As a semi-analytical method.

Turning our attention back to Eq.~\eqref{eq:EigFrameDynSys}, we choose $\Phi_\mm{easy} (t) = \Phi_0 (t)$ [see
Eq.~\eqref{eq:Phi0}]. This leads to $V_\mm{good,I} (t) = + s i \dot{\theta} (t) (\sigma_x + s i \sigma_y)/2$  and $V_\mm{bad,I}
(t) = - s i \dot{\theta} (t) (\sigma_x - s i \sigma_y)/2$. 

Furthermore, following the prescription of Ref.~\cite{roque2021}, we parametrize the control $W(t)$ as  
\begin{equation}
	W (t) = \sum_n W^{(n)} (t) = \sum_n \Delta_\uc^{(n)} (t) \sigma_z + g_\uc^{(n)} (t) \sigma_x,
	\label{eq:W}
\end{equation}
where we represent the control fields using a truncated Fourier series 
\begingroup
\small
\begin{equation}
	\begin{aligned}
		\Delta_\uc^{(j)} (t) &= \! \! \sum_{k=k_\um}^{k_\uM} c_k^{(j)} \left[ 1- \cos\left( 2\pi k \frac{t}{t_\uf} \right)
		\right] +  \sum_{l=l_\um}^{l_\uM} d_l^{(j)} \sin\left( 2\pi l \frac{t}{t_\uf} \right), \\
		g_\uc^{(j)} (t) &= \! \! \sum_{m=m_\um}^{m_\uM} \! \! c_m^{(j)} \left[ 1- \cos\left( 2\pi m \frac{t}{t_\uf} \right)
		\right] \! + \!\! \sum_{n=n_\um}^{n_\uM} \! \! d_n^{(j)} \sin\left( 2\pi n \frac{t}{t_\uf} \right).
	\end{aligned}
	\label{eq:DeltaAndgCorr}
\end{equation}
\endgroup
with $\Delta_\uc^{(j)} (t)$ and $g_\uc^{(j)} (t)$ chosen to vanish at $t=0$ and $t=t_\uf$. 

We can now follow the procedure introduced in Ref.~\cite{roque2021} to find the linear system of equations determining the Fourier
coefficients of Eq.~\eqref{eq:DeltaAndgCorr}, but with the vector of spurious elements (see Eq.~(93) in Ref.~\cite{roque2021} and
Supplemental Material) being given by Eq.~\eqref{eq:EqWI}. Moreover, since we want to preserve the non-reciprocal behavior with
respect to the loop-orientation, we solve simultaneously for the Fourier coefficients that cancel out the effects of $V_\mm{bad,I}
(t)$ for both $s=+1$ and $s=-1$ (see Supplemental Material). This amounts to require that independently of the orientation of the
path we always cancel on average transitions from the gain mode to the lossy mode [see Fig.~\ref{fig:fig03}~(e)].

We compare in Fig.~\ref{fig:fig03}~(c) the average error [see Eq.~\eqref{eq:AvgErr}] between the uncorrected circular control loop
[see Eq.~\eqref{eq:CircPathParam}] and its second-order correction. Simple modifications of the control fields [see
Fig.~\ref{fig:fig03}~(a) and (b)], yielding a modified path in parameter space [see Fig.~\ref{fig:fig03}~(d)], lead to a reduction
of the average error by a few orders of magnitude. Or in other words, it is possible to achieve a non-reciprocal exchange of
energy that is much faster for a comparable error.

\textit{Conclusion ---} 
In conclusion, we have shown how to obtain accurate perturbative solutions describing the evolution of non-Hermitian systems based
on the Magnus expansion. The existence of perturbative solutions further allows us to use the Magnus-based strategy for control to
speed up the non-reciprocal exchange of energy when encircling an exceptional point. Finally, we have introduced two major
modifications to the Magnus-based strategy for control that allow one to (I) deal with problems for which it is hard to solve for
the generator of the ideal dynamics and (II) extend its applicability to non-Hermitian systems.

\begin{appendix}
\clearpage
\thispagestyle{empty}
\onecolumngrid
\begin{center}
\textbf{\large Supplemental Material for: Accelerated Non-Reciprocal Transfer of Energy Around an Exceptional Point}
\end{center}

\section{Perturbation theory for non-Hermitian systems} 
\label{sec:PerturbSols}

In this section, we show in more details how one can use the Magnus expansion to find approximate solutions   
of the dynamial system (Eq.~\eqref{eq:EigFrameDynSysI} of the main text) 
\begin{equation}
	\dot{\Phi}_\uI (t) =
	D_\uI (t) \Phi_\uI (t) =
	\dot{\theta} (t) \left(e^{2 i \Lambda (t)} \sigma_+ - e^{-2 i \Lambda (t)} \sigma_- \right) \Phi_\uI (t).
	\label{eq:DIsup}
\end{equation}

Using the Magnus expansion, we can formally write the exact solution as 
\begin{equation}
	\Phi_\uI (t) = \exp\left[ \sum_{k=1}^\infty \epsilon^k \Omega_k (t) \right] = \mathbbm{1} + \sum_{j=1}^\infty \frac{1}{j!} \left[
	\sum_{k=1}^\infty  \epsilon^k \Omega_k (t) \right]^j,
	\label{eq:MagnusSol}
\end{equation}
where the second equality follows from expanding the exponential function with a Taylor series (Eq.~\eqref{eq:PhiIApprox} of the
main text) and we use the parameter $\epsilon$ for bookkeeping. 

Approximate solutions are found by truncating the series at a desired order in $\epsilon$. Keeping at most terms that are fourth
order in $\epsilon$, we find 
\begin{equation}
	\begin{aligned}
		\Phi_\uI (t) &= \mathbbm{1} + \epsilon \Omega_1 (t) + \epsilon^2\left[ \frac{1}{2} \Omega_1^2 (t) + \Omega_2 (t) \right] +
		\epsilon^3 \left[ \frac{1}{3!} \Omega_1^3 (t) + \frac{1}{2}\left\{\Omega_1 (t),\Omega_2 (t)\right\} + \Omega_3 (t)
		\right] \\
		&\phantom{={}}
		+ \epsilon^4\left[ \frac{1}{4!} \Omega_1^4 (t) + \frac{1}{2} \Omega_2^2 (t) + \frac{1}{2}\left\{\Omega_1
		(t),\Omega_3 (t)\right\} +  \frac{1}{3!} \left\{\Omega_1 (t)^2,\Omega_2 (t)\right\} +  \frac{1}{3!} \Omega_1 (t)
		\Omega_2 (t) \Omega_1 (t) + \Omega_4 (t) \right] \\
		&\phantom{={}}
		+\mathcal{O}\left( \epsilon^5 \right),
	\end{aligned}
	\label{eq:MagnusDysonSol}
\end{equation}
where $\{A_1,A_2\} = A_1 A_2 + A_2 A_1$ denotes the anticommutator of the matrices $A_1$ and $A_2$ and the Magnus elements
$\Omega_k (t)$ with $k\in \{1,4\}$ are given by 
\begin{equation}
	\begin{aligned}
		\Omega_1 (t) &= \int_0^t \di{t_1} D_\uI (t_1) = f^{(1)}_+ (t) \sigma_+ - f^{(1)}_- (t) \sigma_-,\\
		\Omega_2 (t) &=  \frac{1}{2} \int_0^t \di{t_1} \left[ D_\uI (t_1), \Omega_1 (t_1) \right] =  \frac{1}{2} f^{(2)}_z
		(t) \sigma_z, \\
		\Omega_3 (t) &= \int_0^t \di{t_1}\left\{ \frac{1}{2} \left[ D_\uI (t_1), \Omega_2 (t_1) \right] 
		+ \frac{1}{12} \left[ \Omega_1 (t_1), \left[ \Omega_1 (t_1),D_\uI (t_1) \right] \right]  \right\} 
		= f^{(3)}_+ (t) \sigma_+ + f^{(3)}_- (t) \sigma_-, \\
		\Omega_4 (t) &= \int_0^t \di{t_1}\left\{ \frac{1}{2} \left[ D_\uI (t_1), \Omega_3 (t_1) \right] 
			+  \frac{1}{12} \left[ \Omega_2 (t_1), \left[ \Omega_1 (t_1),D_\uI (t_1) \right] \right] 
			+ \frac{1}{12} \left[ \Omega_1 (t_1), \left[ \Omega_2 (t_1),D_\uI (t_1) \right] \right] \right\} \\
			&= \frac{1}{2} f^{(4)}_z (t) \sigma_z,
	\end{aligned}
	\label{eq:MagnusEls4th}
\end{equation}
where  $[A_1,A_2] = A_1 A_2 - A_2 A_1$ denotes the commutator of the matrices $A_1$ and $A_2$. We have decomposed the Magnus
elements in the basis of Pauli matrices and we have introduced $f_\pm^{(2k-1)} (t)$ and $f_z^{(2k)} (t)$ with $k\in \{1,2\}$ to
denote the time-dependent coefficients of the decomposition.  

Substituting Eq.~\eqref{eq:MagnusEls4th} into Eq.~\eqref{eq:MagnusDysonSol} and setting $\epsilon =1$, we find 
\begin{equation}
	\begin{aligned}
		\Phi_\uI^{(4)} (t) &= 
		\left\{ 1 + 
		\frac{1}{2}\left( f_+^{(1)} (t) f_-^{(3)} (t) - f_+^{(1)} (t) f_-^{(1)} (t) - f_-^{(1)} (t)f_+^{(3)} (t)\right) 
		+ \frac{1}{4!} \left[ f_+^{(1)} (t) f_-^{(1)} (t) \right]^2 + \frac{1}{8} \left[ f_z^{(2)} (t)
		\right]^2\right\} \mathbbm{1} \\
		&\phantom{={}} 
		+ \frac{1}{2}\left[ f_z^{(2)} (t) + f_z^{(4)} (t) - \frac{1}{3!} f_-^{(1)} (t) f_+^{(1)} (t) f_z^{(2)} (t)\right]
		\sigma_z 
		+ \left[f_+^{(1)} (t) + f_+^{(3)} (t) - \frac{1}{3!} \left[ f_+^{(1)} (t) \right]^2 f_-^{(1)} (t)
		\right] \sigma_+ \\
		&\phantom{={}}
		- \left[f_-^{(1)} (t) - f_-^{(3)} (t) - \frac{1}{3!} \left[ f_-^{(1)} (t) \right]^2 f_+^{(1)} (t) \right]
		\sigma_-,
	\end{aligned}
	\label{eq:PhiIApprox4th}
\end{equation}
where we used the notation $\Phi_\uI^{(n)} (t)$ introduced in the main text and which is defined via $\Phi_\uI (t) =
\Phi_\uI^{(n)} (t) + \mathcal{O} (\epsilon^{n+1})$. 

Exact closed-form expressions for the coefficients $f_\pm^{(2k-1)} (t)$ and $f_z^{(2k)} (t)$ [see Eq.~\eqref{eq:MagnusEls4th}] are
difficult to obtain. We can, however, find series representations in powers of $1/(\Gamma t_\uf)$ by iteratively integrating by
parts Eq.~\eqref{eq:MagnusEls4th}. The general strategy is reminiscent of the standard procedure used when trying to
approximate the integral of a fast oscillating function multiplied by a slow varying envelope function, but here we need to take into
account that the frequency of the fast oscillating function is explicitly time-dependent. As an example, we show below the first
iteration for the functions $f_\pm^{(1)} (t)$. We have 
\begin{equation}
	\begin{aligned}
		f_\pm^{(1)} (t) = \int_0^t \di{t_1} e^{\pm 2 i \Lambda (t_1)} \dot{\theta} (t_1) 
		&= \int_0^t \di{t_1} \left\{\drv{}{t_1}\left[\mp \frac{i}{2}\frac{1}{\lambda (t_1)} e^{\pm 2 i \Lambda (t_1)}\right] 
		\mp \frac{i}{2}\frac{\dot{\lambda} (t_1)}{\lambda^2 (t_1)} e^{\pm 2 i \Lambda (t_1)} \right\} \dot{\theta} (t_1) \\
		&= \mp \frac{i}{2} \left[ \frac{\dot{\theta} (t)}{\lambda (t)} e^{\pm 2 i \Lambda (t)} - \frac{\dot{\theta} (0)}{\lambda
		(0)}\right] \pm \frac{i}{2} \int_0^t \di{t_1} e^{\pm 2 i \Lambda (t_1)} \left[ \frac{\ddot{\theta} (t_1)}{\lambda (t_1)} -
		\frac{\dot{\lambda} (t1)}{\lambda (t_1)}\frac{\dot{\theta} (t_1)}{\lambda (t_1)}\right]. 
	\end{aligned}
	\label{eq:fpmExact}
\end{equation}

By truncating the series representations at fourth order in  $1/(\Gamma t_\uf)$, we find 
\begin{equation}
	\begin{aligned}
		f_{\pm}^{(1)} (t) &= \mp \frac{i}{2} \left[ \frac{\dot{\theta} (t)}{\lambda (t)} e^{\pm 2 i \Lambda (t)} 
		- \frac{\dot{\theta} (0)}{\lambda(0)}\right]
		+\frac{1}{4}\left[ \left(\frac{\ddot{\theta} (t)}{\lambda^2 (t)}
		-\frac{\dot{\theta}(t)}{\lambda (t)} \frac{\dot{\lambda}(t)}{\lambda^2 (t)}\right)e^{\pm 2 i \Lambda (t)} 
		- \left(\frac{\ddot{\theta} (0)}{\lambda^2 (0)}
		-\frac{\dot{\theta}(0)}{\lambda (0)} \frac{\dot{\lambda} (0)}{\lambda^2 (0)}\right) \right]\\
		&\phantom{={}} 
		\pm \frac{i}{8} \left\{\left[ \frac{\theta^{(3)}(t)}{\lambda^3 (t)} 
		-\frac{3\ddot{\theta}(t) \dot{\lambda} (t)}{\lambda^4 (t)} 
		+ \frac{\dot{\theta}(t)}{\lambda (t)}\left(\frac{3 \dot{\lambda}^2(t)}{\lambda^4 (t)} -
		\frac{\ddot{\lambda}(t)}{\lambda^3 (t)} \right)\right] e^{\pm 2 i \Lambda (t)}
		- \left[\frac{\theta^{(3)}(0)}{\lambda^3 (0)} 
		-\frac{3\ddot{\theta}(0) \dot{\lambda} (0)}{\lambda^4 (0)} 
		+ \frac{\dot{\theta}(0)}{\lambda (0)}\left( \frac{3\dot{\lambda}^2(0)}{\lambda^4 (0)} -
		\frac{\ddot{\lambda}(0)}{\lambda^3 (0)} \right)\right]
		\right\} \\
		&\phantom{={}} 
		-\frac{1}{16} \left\{\left[\frac{\theta^{(4)}(t)}{\lambda^4 (t)}
		-\frac{6\theta^{(3)}(t) \dot{\lambda}(t)}{\lambda^5 (t)}
		+ \frac{\ddot{\theta}(t)}{\lambda^2 (t)}\left( \frac{15 \dot{\lambda}^2 (t)}{\lambda^4 (t)} -
		\frac{4\ddot{\lambda}(t)}{\lambda^3 (t)}\right)
		+ \frac{\dot{\theta}(t)}{\lambda (t)} \left( \frac{10 \dot{\lambda} (t) \ddot{\lambda}(t)}{\lambda^5 (t)} 
		- \frac{\lambda^{(3)}(t)}{\lambda^4 (t)} -  \frac{15\dot{\lambda}^3 (t)}{\lambda^6 (t)}\right)
		\right]e^{\pm 2 i \Lambda (t)} \right. \\
		&\phantom{={}-\frac{1}{16}[} \left.
		-\left[\frac{\theta^{(4)}(0)}{\lambda^4 (0)}
		-\frac{6\theta^{(3)}(0) \dot{\lambda}(0)}{\lambda^5 (0)}
		+ \frac{\ddot{\theta}(0)}{\lambda^2 (0)}\left( \frac{15 \dot{\lambda}^2 (0)}{\lambda^4 (0)} -
		\frac{4\ddot{\lambda}(0)}{\lambda^3 (0)}\right)
		+ \frac{\dot{\theta}(0)}{\lambda (0)} \left( \frac{10 \dot{\lambda} (0) \ddot{\lambda}(0)}{\lambda^5 (0)} 
		- \frac{\lambda^{(3)}(0)}{\lambda^4 (0)} - \frac{15\dot{\lambda}^3 (0)}{\lambda^6 (0)}\right)
		\right] \right\} \\
		&\phantom{={}}
		+\mathcal{O}\left[ \frac{1}{\left( \Gamma t_\uf \right)^5} \right],
	\end{aligned}
	\label{eq:fpm1Approx}
\end{equation}
where we have defined $h^{(n)} (t) = \drv{^n h (t)}{t^n}$ for $n>2$. Proceeding similarly, we find 
\begin{equation}
	\begin{aligned}
		f_z^{(2)} (t) &= -i \int_0^t \di{t_1} \left ( \frac{\dot{\theta}^2 (t_1)}{\lambda (t_1)} \right)
		+ \frac{1}{4}\frac{\dot{\theta}(0)}{\lambda(0)}\frac{\dot{\theta}(t)}{\lambda(t)}\left(e^{2i \Lambda(t)} - e^{-2i \Lambda(t)}\right) 
		+ \frac{i}{4} \int_0^t \di{t_1} \left[\frac{\dot{\theta} (t_1)}{\lambda (t_1)}\drv{}{t_1}\left( \frac{\ddot{\theta}
		(t_1)}{\lambda^2 (t_1)} - \frac{\dot{\theta} (t_1) \dot{\lambda} (t_1)}{\lambda^3 (t_1)} \right) \right]\\
		&\phantom{={}} 
		+ \frac{i}{8}\left[
		\left(\frac{\dot{\theta} (0) \dot{\lambda} (0)}{\lambda^3 (0)}-\frac{\ddot{\theta}
		(0)}{\lambda^2 (0)}\right) \frac{\dot{\theta} (t) }{\lambda (t)} \left(e^{2 i \Lambda(t)} + e^{-2 i \Lambda(t)}\right)
		- \frac{\dot{\theta} (0) }{\lambda (0)} \left(\frac{\dot{\theta} (t) \dot{\lambda} (t)}{\lambda^3(t)} 
		- \frac{\ddot{\theta}(t)}{\lambda^2 (t)}\right) \left(e^{2 i \Lambda(t)} + e^{-2 i \Lambda(t)}\right)
		\right] \\
		&\phantom{={}} 
		%
		-\frac{1}{16}\left\{
		\left[
		\frac{\theta^{(3)}(0)}{\lambda^3 (0)} 
		-\frac{3 \ddot{\theta}(0) \dot{\lambda}(0)}{\lambda^4 (0)} +\frac{\dot{\theta} (0)}{\lambda (0)} 
		\left(\frac{3 \dot{\lambda}^2(0)}{\lambda^4 (0)} - \frac{\ddot{\lambda}(0)}{\lambda^3 (0)}\right)
		\right]
		\frac{\dot{\theta} (t)}{\lambda (t)} 
		+ 3\frac{\dot{\theta} (0)}{\lambda (0)} \left(\frac{\ddot{\theta}(t) \dot{\lambda}(t)}{\lambda^4 (t)}
		-\frac{\dot{\theta} (t) \dot{\lambda}^2 (t)}{\lambda^5 (t)}\right) \right .\\
		&\phantom{={}-\frac{1}{16}\{} \left .
		+ \left(\frac{\ddot{\theta} (0)}{\lambda^2 (0)}-\frac{\dot{\theta} (0) \dot{\lambda} (0)}{\lambda^3 (0)}\right)
		\left(\frac{\ddot{\theta} (t)}{\lambda^2 (t)}-\frac{\dot{\theta} (t) \dot{\lambda} (t)}{\lambda^3 (t)}\right) 
		+ \frac{\dot{\theta} (0)}{\lambda (0)} \left(\frac{\dot{\theta}(t) \ddot{\lambda} (t)}{\lambda^4 (t)}
		-\frac{\theta^{(3)}(t)}{\lambda^3 (t)}\right) \right\} \left(e^{2 i \Lambda (t)}-e^{-2 i \Lambda (t)}\right) \\
		&\phantom{={}} 
		+\mathcal{O}\left[ \frac{1}{\left( \Gamma t_\uf \right)^5} \right],
	\end{aligned}
	\label{eq:fz2Approx}
\end{equation}

\begin{equation}
	\begin{aligned}
		f_\pm^{(3)} (t) &= \pm \frac{1}{4} \left( \frac{\dot{\theta} (t)}{\lambda (t)} e^{\pm 2 i \Lambda (t)} +
		\frac{\dot{\theta} (0)}{\lambda (0)}\right) \int_0^t \di{t_1}  \left(\frac{\dot{\theta}^2 (t_1)}{\lambda (t_1)}
		\right) \\
		&\phantom{={}}
		+\frac{i}{2}\left\{ -\frac{1}{12}\frac{\dot{\theta} (0)}{\lambda(0)} \int_0^t \di{t_1} \left(
		\frac{\dot{\theta}(t_1)}{\lambda(t_1)} \drv{}{t_1}\frac{\dot{\theta} (t_1)}{\lambda (t_1)}\right)
		+\frac{1}{24} \frac{\dot{\theta}^2 (0) \dot{\theta} (t)}{\lambda^2 (0)\lambda (t)}
		\left(e^{\pm 2 i \Lambda(t)}-e^{\mp 2 i \Lambda(t)}\right) \right .\\
		&\phantom{={} + \frac{i}{2} \Big[}
		+\frac{1}{3} \left( \frac{\dot{\theta}^3 (t)}{\lambda^3 (t)}e^{\pm 2 i \Lambda(t)} -\frac{\dot{\theta}^3(0)}{\lambda^3 (0)}\right)
		+\frac{1}{24} \frac{\dot{\theta} (0)}{\lambda (0)}\left(\frac{\dot{\theta}^2 (t)}{\lambda^2 (t)} e^{\pm 4
		i\Lambda(t)} - \frac{\dot{\theta}^2(0)}{\lambda^2 (0)} \right)\\
		&\phantom{={} + \frac{i}{2} \Big[} \left .
		+\frac{1}{4} \left[ 
		\left(\frac{\ddot{\theta} (t)}{\lambda^2 (t)}-\frac{\dot{\theta} (t) \dot{\lambda} (t)}{\lambda^3 (t)} \right) 
		e^{\pm 2 i \Lambda (t)} 
		+ \left(\frac{\ddot{\theta} (0)}{\lambda^2 (0)}-\frac{\dot{\theta} (0) \dot{\lambda}(0)}{\lambda^3 (0)}\right) \right] 
		\int_0^t \di{t_1} \left( \frac{\dot{\theta}^2 (t_1)}{\lambda (t_1)} \right) \right\} \\
		&\phantom{={}} \pm \frac{1}{16}\left\{
		\int_0^t \di{t_1} \left[\frac{\dot{\theta} (0) \dot{\theta} (t_1)}{\lambda (0) \lambda (t_1)} \left(3
		\frac{\dot{\lambda} (t_1)}{\lambda^2 (t_1)}\drv{}{t_1} \frac{\dot{\theta} (t_1)}{\lambda (t_1)} +
		\frac{\dot{\theta} (t_1) \ddot{\lambda} (t_1)}{\lambda^3 (t_1)} - \frac{\theta^{(3)} (t_1)}{\lambda^2 (t_1)}
		\right) \right. \right .\\
		&\phantom{= \pm \frac{1}{16} \{\int_0^t \di{t_1}\Big[+}
		+ \frac{\dot{\theta}^2 (t_1)}{\lambda (0) \lambda (t_1)} \left( 3 \frac{\dot{\lambda} (0)}{\lambda^2 (0)}
		\left[\drv{}{t_1} \frac{\dot{\theta} (t_1)}{\lambda (t_1)}\right]_{t_1 = 0} + \frac{\dot{\theta} (0)
		\ddot{\lambda} (0)}{\lambda^3 (0)} - \frac{\theta^{(3)} (0)}{\lambda^2 (0)}  \right) \\
		&\phantom{= \pm \frac{1}{16} \{\int_0^t \di{t_1}\Big[+}
		\left .
		+ \frac{1}{3} \left( \frac{\ddot{\theta} (0) \dot{\theta} (t_1)}{\lambda^2 (0) \lambda (t_1)} -
		\frac{\dot{\theta} (0) \dot{\lambda} (0) \dot{\theta} (t_1)}{\lambda^3 (0) \lambda (t_1)}\right)\drv{}{t_1}
		\frac{\dot{\theta} (t_1)}{\lambda (t_1)}\right] \\
		&\phantom{=\pm \frac{1}{16}[ }
		- \int_0^t \di{t_1} \left[\frac{\dot{\theta} (t_1)}{\dot{\lambda} (t_1)}
		\left( -3 \frac{\dot{\lambda}(t_1)}{\lambda^2 (t_1)} \drv{}{t_1}\frac{\dot{\theta}(t_1)}{\lambda (t_1)}
		- \frac{\dot{\theta} (t_1) \ddot{\lambda}(t_1)}{\lambda^3 (t_1)} + \frac{\theta^{(3)} (t_1)}{\lambda^2 (t_1)}  \right) \right] 
		\frac{\dot{\theta} (t)}{\lambda (t)} e^{\pm 2 i \Lambda(t)} \\
		&\phantom{=\pm \frac{1}{16}[ }
		+\frac{1}{3} \left[ \frac{\ddot{\theta} (0) \dot{\theta} (t)}{\lambda^2 (0)} - \frac{\dot{\theta} (0)}{\lambda (0)}
		\left(\frac{\dot{\lambda} (0) \dot{\theta} (t)}{\lambda^2 (0)} + \frac{1}{2} \drv{}{t} \frac{\dot{\theta} (t)}{\lambda (t)}\right)
		\right] \frac{\dot{\theta} (0)}{\lambda (0) \lambda (t)} e^{\mp 2 i \Lambda(t)} \\
		&\phantom{=\pm \frac{1}{16}[ }
		+\frac{1}{3}  \left[ \frac{\ddot{\theta} (0) \dot{\theta} (t)}{2 \lambda^2 (0)} - \frac{\dot{\theta} (0)}{\lambda(0)}
		\left(\frac{\dot{\lambda} (0) \dot{\theta} (t)}{2 \lambda^2 (0)} + \drv{}{t} \frac{\dot{\theta} (t)}{\lambda (t)}\right) \right]
		\frac{\dot{\theta} (t)}{\lambda^2 (t)}e^{\pm 4 i \Lambda(t)} \\
		&\phantom{=\pm \frac{1}{16}[ }
		- \left[\left(\frac{\dot{\theta}^2 (0)}{6 \lambda^2 (0) \lambda (t)} + \frac{16}{3} \frac{\dot{\theta}^2
		(t)}{\lambda^3 (t)}\right) \drv{}{t} \frac{\dot{\theta} (t)}{\lambda (t)} - \int_0^t \di{t_1} \frac{\dot{\theta}^2 (t_1)}{\lambda (t_1)} 
		\left(3 \frac{\dot{\lambda} (t)}{\lambda^3 (t)} \drv{}{t}  \frac{\dot{\theta} (t)}{\lambda (t)} 
		+ \frac{\dot{\theta} (t)\ddot{\lambda} (t)}{\lambda^4 (t)} - \frac{\theta^{(3)} (t)}{\lambda^3 (t)} \right)\right]
		e^{\pm 2 i \Lambda(t)} \\
		&\phantom{=\pm \frac{1}{16}[ } \left .
		+\frac{11}{2} \frac{\dot{\theta}^2 (0)}{\lambda^3 (0)}\left[\drv{}{t} \frac{\dot{\theta} (t)}{\lambda (t)}\right]_{t=0} \right\}
		+ \mathcal{O}\left[ \frac{1}{\left( \Gamma t_\uf \right)^5} \right],
	\end{aligned}
	\label{eq:fpm3Approx}
\end{equation}
where $[\ud f(t)/\ud t]_{t=0} = \dot{f} (0)$ denotes that we evaluate the derivative at $t=0$. Finally, we have
\begin{equation}
	\begin{aligned}
		f_z^{(4)} (t) &= \frac{i}{3}\left[ \int_0^t \di{t_1} \frac{\dot{\theta} (t_1)}{\lambda (t_1)} \left(
		\frac{\dot{\theta}^2 (0) \dot{\theta} (t_1) }{4 \lambda^2 (0)} + \frac{\dot{\theta}^3 (t_1)}{\lambda^2 (t_1)}
		+ \frac{1}{2} \left(\drv{}{t_1} \frac{\dot{\theta} (t_1)}{\lambda(t_1)}\right)\int_0^{t_1} \di{t_2}
		\frac{\dot{\theta}^2 (t_2)}{\lambda (t_2)} \right) \right. \\
		&\phantom{=\frac{i}{3} [} \left.
		+ \frac{1}{2} \frac{\dot{\theta} (0) \dot{\theta} (t)}{\lambda (0) \lambda (t)} \int_0^t \di{t_1}
		\left(\frac{\dot{\theta}^2 (t_1)}{\lambda (t_1)}\right) \left( e^{2 i \Lambda(t)}  +  e^{-2 i
		\Lambda(t)}\right)\right] \\ 
		&\phantom{={}}
		+\frac{1}{4}\left\{ 
		\frac{1}{4} \frac{\dot{\theta} (0) \dot{\theta} (t)}{\lambda (0) \lambda (t)}\left(- \frac{1}{3} \int_0^t \di{t_1} 
		\left( \frac{\dot{\theta} (t_1)}{\lambda (t_1)} \drv{}{t_1} \frac{\dot{\theta} (t_1)}{\lambda (t_1)} \right) +
		\frac{\dot{\theta}^2 (t)}{6 \lambda^2 (t)} - \frac{3 \dot{\theta}^2 (0)}{2 \lambda^2 (0)}\right) 
		\right. \\
		&\phantom{={} +\frac{1}{4}\Big[}
		+ \frac{\dot{\theta} (t)}{3 \lambda (t) \lambda (0)} \left[\drv{}{t}\frac{\dot{\theta} (t)}{\lambda (t)}
		\right]_{t=0} \int_0^t \di{t_1} \left( \frac{\dot{\theta}^2 (t_1)}{\lambda (t_1)} \right) \\
		&\phantom{={} +\frac{1}{4}\Big[} \left .
		- \frac{\dot{\theta} (0)}{3 \lambda (0)} \left[ \frac{\dot{\theta}^3 (t)}{\lambda^3 (t)} + \frac{1}{\lambda (t)}
		\left(\drv{}{t}\frac{\dot{\theta} (t)}{\lambda (t)}\right) \int_0^t \di{t_1} \left( \frac{\dot{\theta}^2 (t_1)}{\lambda (t_1)} 
		\right) \right]\right\} \left(  e^{2 i \Lambda(t)} - e^{-2 i \Lambda(t)}\right)\\
		&\phantom{={}}
		-\frac{\dot{\theta}^2 (0) \dot{\theta}^2 (t)}{96 \lambda^2 (0) \lambda^2 (t)} \left(  e^{4 i \Lambda(t)} - e^{-4 i \Lambda(t)}\right)
		+ \mathcal{O}\left[ \frac{1}{\left( \Gamma t_\uf \right)^5} \right].
	\end{aligned}
	\label{eq:fz4Approx}
\end{equation}

Substituting the truncated series representations of $f^{(2k-1)}_\pm (t)$ and $f^{(2k)}_z (t)$ $[k\in\{1,2\}]$ in
Eq.~\eqref{eq:PhiIApprox4th}, we can evaluate the matrix elements $\cv_i^\mathsf{T} \Phi (t) \cv_j$ with $i,\,j\in{\uG,\uL}$ and
their modulus squared (not shown here due to the length of the expression). 

The asymptotic expression shown in the main text [see Eq.~\eqref{eq:ApproxSolEls}] is obtained by keeping solely the exponential
large terms that governed the dynamics in the long-time regime, i.e., $\Gamma t \gg 1$.

\section{Magnus-based strategy for control}
\label{sec:Magnus}

In this section, we show in more detail how we obtained the linear system of equations determining the Fourier coefficients of the
control fields $\Delta_\uc (t)$ and $g_\uc (t)$. 

In the interaction picture defined by $\Phi_0 (t)$ (see Eq.~\eqref{eq:Phi0} of the main text), the control matrix $W (t)$ (see
Eq.~\eqref{eq:W} of the main text) takes the form 
\begin{equation}
	\begin{aligned}
		W_\uI (t) &= \sum_n \tilde{w}_z^{(n)} (t) \sigma_z + \tilde{w}_x^{(n)} (t) \sigma_x + \tilde{w}_y^{(n)} (t) \sigma_y \\
		&=\sum_n \left[-\left( \frac{i \frac{\Gamma}{2} + \Delta (t)}{\lambda (t)} \Delta_\uc^{(n)} (t) 
		- \frac{g (t)}{\lambda (t)} g_\uc^{(n)} (t) \right) \sigma_z \right.
		- \cos\left[ \Lambda (t) \right] \left(  \frac{g (t)}{\lambda (t)} \Delta_\uc^{(n)} (t) 
		+ \frac{i \frac{\Gamma}{2} + \Delta (t)}{\lambda (t)} g_\uc^{(n)} (t) \right) \sigma_x \\
		&\phantom{={}} \left.
		+ \sin\left[ \Lambda (t) \right] \left(\frac{g (t)}{\lambda (t)} \Delta_\uc^{(n)} (t) 
		+ \frac{i \frac{\Gamma}{2} + \Delta (t)}{\lambda (t)} g_\uc^{(n)} (t) \right)\sigma_y \right].
	\end{aligned}
	\label{eq:WILin}
\end{equation}

The first order correction is found by solving  Eq.~\eqref{eq:EqWI} of the main text for $n=1$. Since we want $W(t)$ to cancel the
effects of $V_\mm{bad} (t)$ independently of the orientation of the control loop, we must solve the system of equations 
\begin{equation}
	\begin{aligned}
		\int_0^{t_\uf} \di{t} W_\uI^{(1)} (t) &= -i \int_0^{t_\uf} \di{t} V_{\mm{bad},\uI}^{\circlearrowright} (t), \\
		\int_0^{t_\uf} \di{t} W_\uI^{(1)} (t) &= -i \int_0^{t_\uf} \di{t} V_{\mm{bad},\uI}^{\circlearrowleft} (t).
	\end{aligned}
	\label{eq:LinSysStep1}
\end{equation}
where we have defined 
\begin{equation}
	\begin{aligned}
		V_{\mm{bad},\uI}^{\circlearrowright} (t) &= \tilde{v}_z^\circlearrowright (t) \sigma_z +
		\tilde{v}_x^\circlearrowright (t) \sigma_x + \tilde{v}_y^\circlearrowright (t) \sigma_y = e^{2 i \Lambda (t)]}
		\dot{\theta} (t) \left( \sigma_x + i \sigma_y \right), \\
		V_{\mm{bad},\uI}^{\circlearrowleft} (t) &= \tilde{v}_z^\circlearrowleft (t) \sigma_z +
		\tilde{v}_x^\circlearrowleft (t) \sigma_x + \tilde{v}_y^\circlearrowleft (t) \sigma_y = -e^{-2 i \Lambda (t)]}
		\dot{\theta} (t) \left( \sigma_x - i \sigma_y \right).
	\end{aligned}
	\label{eq:VIPauli}
\end{equation}

Using the decomposition of $W_\uI (t)$ [see Eq.~\eqref{eq:WILin}] and $V_{\mm{bad},\uI}^s (t)$ [see Eq.~\eqref{eq:VIPauli}] into
the basis of Pauli matrices and taking into account that the coefficients of the decomposition are complex,
Eq.~\eqref{eq:LinSysStep1} can be written as 
\begin{equation}
	\begin{aligned}
		\mm{Re}\left[\int_0^{t_\uf} \di{t} \tilde{w}_z^{(1)}\right] &= \mm{Re}\left[-i \int_0^{t_\uf} \di{t} \tilde{v}_z^\circlearrowright (t)\right], \\
		\mm{Im}\left[\int_0^{t_\uf} \di{t} \tilde{w}_z^{(1)}\right] &= \mm{Im}\left[-i \int_0^{t_\uf} \di{t} \tilde{v}_z^\circlearrowright (t)\right], \\
		\mm{Re}\left[\int_0^{t_\uf} \di{t} \tilde{w}_x^{(1)}\right] &= \mm{Re}\left[-i \int_0^{t_\uf} \di{t} \tilde{v}_x^\circlearrowright (t)\right], \\
		\mm{Im}\left[\int_0^{t_\uf} \di{t} \tilde{w}_x^{(1)}\right] &= \mm{Im}\left[-i \int_0^{t_\uf} \di{t} \tilde{v}_x^\circlearrowright (t)\right], \\
		\mm{Re}\left[\int_0^{t_\uf} \di{t} \tilde{w}_y^{(1)}\right] &= \mm{Re}\left[-i \int_0^{t_\uf} \di{t} \tilde{v}_y^\circlearrowright (t)\right], \\
		\mm{Im}\left[\int_0^{t_\uf} \di{t} \tilde{w}_y^{(1)}\right] &= \mm{Im}\left[-i \int_0^{t_\uf} \di{t} \tilde{v}_y^\circlearrowright (t)\right], \\
		\mm{Re}\left[\int_0^{t_\uf} \di{t} \tilde{w}_z^{(1)}\right] &= \mm{Re}\left[-i \int_0^{t_\uf} \di{t} \tilde{v}_z^\circlearrowleft (t)\right], \\
		\mm{Im}\left[\int_0^{t_\uf} \di{t} \tilde{w}_z^{(1)}\right] &= \mm{Im}\left[-i \int_0^{t_\uf} \di{t} \tilde{v}_z^\circlearrowleft (t)\right], \\
		\mm{Re}\left[\int_0^{t_\uf} \di{t} \tilde{w}_x^{(1)}\right] &= \mm{Re}\left[-i \int_0^{t_\uf} \di{t} \tilde{v}_x^\circlearrowleft (t)\right], \\
		\mm{Im}\left[\int_0^{t_\uf} \di{t} \tilde{w}_x^{(1)}\right] &= \mm{Im}\left[-i \int_0^{t_\uf} \di{t} \tilde{v}_x^\circlearrowleft (t)\right], \\
		\mm{Re}\left[\int_0^{t_\uf} \di{t} \tilde{w}_y^{(1)}\right] &= \mm{Re}\left[-i \int_0^{t_\uf} \di{t} \tilde{v}_y^\circlearrowleft (t)\right], \\
		\mm{Im}\left[\int_0^{t_\uf} \di{t} \tilde{w}_y^{(1)}\right] &= \mm{Im}\left[-i \int_0^{t_\uf} \di{t} \tilde{v}_y^\circlearrowleft (t)\right],
	\end{aligned}
	\label{eq:LinSysStep2}
\end{equation}

Substituting Eq.~\eqref{eq:DeltaAndgCorr} of the main text into Eq.~\eqref{eq:LinSysStep2}, we can carry out the time integration
and we are left with a linear system of $12$ equations for the unknown Fourier coefficients. As shown in Ref.~\cite{roque2021}, the
system of equations can be written in matrix form as
\begin{equation}
	M \xv^{(1)} = \yv^{(1)}
	\label{eq:LinSysMatrix}
\end{equation}
with $M$ a known $12 \times N_\mm{coeffs}$ matrix characterizing the evolution of the system under the flow $\Phi_0 (t)$,
$\yv^{(1)}$ is the known vector of length $12$ that encodes the spurious elements, and $\xv^{(1)}$ is the unknown vector of Fourier
coefficients of length $N_\mm{coeffs}$. Here, $N_\mm{coeffs}$ is the total number of Fourier coefficients that one is free to
choose. As noted in Ref.~\cite{roque2021}, for $N_\mm{coeffs} \neq 12$ the system of equations can be solved using the
Moore-Penrose pseudo-inverse. 

Higher-order coefficients are found by solving the linear system of equations 
\begin{equation}
	M \xv^{(n)} = \yv^{(n)},
	\label{eq:LinSysMatrix2}
\end{equation}
where $M$ is the same matrix as in Eq.~\eqref{eq:LinSysMatrix} and the vector of spurious elements $\yv^{(n)}$ is determined using
Eq.~\eqref{eq:EqWI} of the main text. 

\end{appendix}


\begin{thebibliography}{37}%
\makeatletter
\providecommand \@ifxundefined [1]{%
 \@ifx{#1\undefined}
}%
\providecommand \@ifnum [1]{%
 \ifnum #1\expandafter \@firstoftwo
 \else \expandafter \@secondoftwo
 \fi
}%
\providecommand \@ifx [1]{%
 \ifx #1\expandafter \@firstoftwo
 \else \expandafter \@secondoftwo
 \fi
}%
\providecommand \natexlab [1]{#1}%
\providecommand \enquote  [1]{``#1''}%
\providecommand \bibnamefont  [1]{#1}%
\providecommand \bibfnamefont [1]{#1}%
\providecommand \citenamefont [1]{#1}%
\providecommand \href@noop [0]{\@secondoftwo}%
\providecommand \href [0]{\begingroup \@sanitize@url \@href}%
\providecommand \@href[1]{\@@startlink{#1}\@@href}%
\providecommand \@@href[1]{\endgroup#1\@@endlink}%
\providecommand \@sanitize@url [0]{\catcode `\\12\catcode `\$12\catcode
  `\&12\catcode `\#12\catcode `\^12\catcode `\_12\catcode `\%12\relax}%
\providecommand \@@startlink[1]{}%
\providecommand \@@endlink[0]{}%
\providecommand \url  [0]{\begingroup\@sanitize@url \@url }%
\providecommand \@url [1]{\endgroup\@href {#1}{\urlprefix }}%
\providecommand \urlprefix  [0]{URL }%
\providecommand \Eprint [0]{\href }%
\providecommand \doibase [0]{https://doi.org/}%
\providecommand \selectlanguage [0]{\@gobble}%
\providecommand \bibinfo  [0]{\@secondoftwo}%
\providecommand \bibfield  [0]{\@secondoftwo}%
\providecommand \translation [1]{[#1]}%
\providecommand \BibitemOpen [0]{}%
\providecommand \bibitemStop [0]{}%
\providecommand \bibitemNoStop [0]{.\EOS\space}%
\providecommand \EOS [0]{\spacefactor3000\relax}%
\providecommand \BibitemShut  [1]{\csname bibitem#1\endcsname}%
\let\auto@bib@innerbib\@empty
\bibitem [{\citenamefont {El-Ganainy}\ \emph {et~al.}(2018)\citenamefont
  {El-Ganainy}, \citenamefont {Makris}, \citenamefont {Khajavikhan},
  \citenamefont {Musslimani}, \citenamefont {Rotter},\ and\ \citenamefont
  {Christodoulides}}]{el-ganainy2018}%
  \BibitemOpen
  \bibfield  {author} {\bibinfo {author} {\bibfnamefont {R.}~\bibnamefont
  {El-Ganainy}}, \bibinfo {author} {\bibfnamefont {K.~G.}\ \bibnamefont
  {Makris}}, \bibinfo {author} {\bibfnamefont {M.}~\bibnamefont {Khajavikhan}},
  \bibinfo {author} {\bibfnamefont {Z.~H.}\ \bibnamefont {Musslimani}},
  \bibinfo {author} {\bibfnamefont {S.}~\bibnamefont {Rotter}},\ and\ \bibinfo
  {author} {\bibfnamefont {D.~N.}\ \bibnamefont {Christodoulides}},\ }\bibfield
   {title} {\bibinfo {title} {Non-hermitian physics and pt symmetry},\ }\href
  {https://doi.org/10.1038/nphys4323} {\bibfield  {journal} {\bibinfo
  {journal} {Nature Physics}\ }\textbf {\bibinfo {volume} {14}},\ \bibinfo
  {pages} {11} (\bibinfo {year} {2018})}\BibitemShut {NoStop}%
\bibitem [{\citenamefont {Kato}(1995)}]{kato1995}%
  \BibitemOpen
  \bibfield  {author} {\bibinfo {author} {\bibfnamefont {T.}~\bibnamefont
  {Kato}},\ }\href@noop {} {\emph {\bibinfo {title} {{Perturbation Theory for
  Linear Operators}}}},\ \bibinfo {edition} {2nd}\ ed.,\ Classics in
  Mathematics\ (\bibinfo  {publisher} {Springer, Berlin, Heidelberg},\ \bibinfo
  {year} {1995})\BibitemShut {NoStop}%
\bibitem [{\citenamefont {Berry}(2004)}]{berry2004}%
  \BibitemOpen
  \bibfield  {author} {\bibinfo {author} {\bibfnamefont {M.~V.}\ \bibnamefont
  {Berry}},\ }\bibfield  {title} {\bibinfo {title} {Physics of nonhermitian
  degeneracies},\ }\href {https://doi.org/10.1023/B:CJOP.0000044002.05657.04}
  {\bibfield  {journal} {\bibinfo  {journal} {Czechoslovak Journal of Physics}\
  }\textbf {\bibinfo {volume} {54}},\ \bibinfo {pages} {1039} (\bibinfo {year}
  {2004})}\BibitemShut {NoStop}%
\bibitem [{\citenamefont {Seyranian}\ \emph {et~al.}(2005)\citenamefont
  {Seyranian}, \citenamefont {Kirillov},\ and\ \citenamefont
  {Mailybaev}}]{seyranian2005}%
  \BibitemOpen
  \bibfield  {author} {\bibinfo {author} {\bibfnamefont {A.~P.}\ \bibnamefont
  {Seyranian}}, \bibinfo {author} {\bibfnamefont {O.~N.}\ \bibnamefont
  {Kirillov}},\ and\ \bibinfo {author} {\bibfnamefont {A.~A.}\ \bibnamefont
  {Mailybaev}},\ }\bibfield  {title} {\bibinfo {title} {Coupling of eigenvalues
  of complex matrices at diabolic and exceptional points},\ }\href
  {https://doi.org/10.1088/0305-4470/38/8/009} {\bibfield  {journal} {\bibinfo
  {journal} {Journal of Physics A: Mathematical and General}\ }\textbf
  {\bibinfo {volume} {38}},\ \bibinfo {pages} {1723} (\bibinfo {year}
  {2005})}\BibitemShut {NoStop}%
\bibitem [{\citenamefont {Heiss}(2012)}]{heiss2012}%
  \BibitemOpen
  \bibfield  {author} {\bibinfo {author} {\bibfnamefont {W.~D.}\ \bibnamefont
  {Heiss}},\ }\bibfield  {title} {\bibinfo {title} {The physics of exceptional
  points},\ }\href {https://doi.org/10.1088/1751-8113/45/44/444016} {\bibfield
  {journal} {\bibinfo  {journal} {Journal of Physics A: Mathematical and
  Theoretical}\ }\textbf {\bibinfo {volume} {45}},\ \bibinfo {pages} {444016}
  (\bibinfo {year} {2012})}\BibitemShut {NoStop}%
\bibitem [{\citenamefont {Lin}\ \emph {et~al.}(2011)\citenamefont {Lin},
  \citenamefont {Ramezani}, \citenamefont {Eichelkraut}, \citenamefont
  {Kottos}, \citenamefont {Cao},\ and\ \citenamefont
  {Christodoulides}}]{lin2011}%
  \BibitemOpen
  \bibfield  {author} {\bibinfo {author} {\bibfnamefont {Z.}~\bibnamefont
  {Lin}}, \bibinfo {author} {\bibfnamefont {H.}~\bibnamefont {Ramezani}},
  \bibinfo {author} {\bibfnamefont {T.}~\bibnamefont {Eichelkraut}}, \bibinfo
  {author} {\bibfnamefont {T.}~\bibnamefont {Kottos}}, \bibinfo {author}
  {\bibfnamefont {H.}~\bibnamefont {Cao}},\ and\ \bibinfo {author}
  {\bibfnamefont {D.~N.}\ \bibnamefont {Christodoulides}},\ }\bibfield  {title}
  {\bibinfo {title} {Unidirectional invisibility induced by
  $\mathcal{P}\mathcal{T}$-symmetric periodic structures},\ }\href
  {https://doi.org/10.1103/PhysRevLett.106.213901} {\bibfield  {journal}
  {\bibinfo  {journal} {Phys. Rev. Lett.}\ }\textbf {\bibinfo {volume} {106}},\
  \bibinfo {pages} {213901} (\bibinfo {year} {2011})}\BibitemShut {NoStop}%
\bibitem [{\citenamefont {Regensburger}\ \emph {et~al.}(2012)\citenamefont
  {Regensburger}, \citenamefont {Bersch}, \citenamefont {Miri}, \citenamefont
  {Onishchukov}, \citenamefont {Christodoulides},\ and\ \citenamefont
  {Peschel}}]{regensburger2012}%
  \BibitemOpen
  \bibfield  {author} {\bibinfo {author} {\bibfnamefont {A.}~\bibnamefont
  {Regensburger}}, \bibinfo {author} {\bibfnamefont {C.}~\bibnamefont
  {Bersch}}, \bibinfo {author} {\bibfnamefont {M.-A.}\ \bibnamefont {Miri}},
  \bibinfo {author} {\bibfnamefont {G.}~\bibnamefont {Onishchukov}}, \bibinfo
  {author} {\bibfnamefont {D.~N.}\ \bibnamefont {Christodoulides}},\ and\
  \bibinfo {author} {\bibfnamefont {U.}~\bibnamefont {Peschel}},\ }\bibfield
  {title} {\bibinfo {title} {Parity-time synthetic photonic lattices},\ }\href
  {https://doi.org/10.1038/nature11298} {\bibfield  {journal} {\bibinfo
  {journal} {Nature}\ }\textbf {\bibinfo {volume} {488}},\ \bibinfo {pages}
  {167} (\bibinfo {year} {2012})}\BibitemShut {NoStop}%
\bibitem [{\citenamefont {Feng}\ \emph {et~al.}(2013)\citenamefont {Feng},
  \citenamefont {Xu}, \citenamefont {Fegadolli}, \citenamefont {Lu},
  \citenamefont {Oliveira}, \citenamefont {Almeida}, \citenamefont {Chen},\
  and\ \citenamefont {Scherer}}]{feng2013}%
  \BibitemOpen
  \bibfield  {author} {\bibinfo {author} {\bibfnamefont {L.}~\bibnamefont
  {Feng}}, \bibinfo {author} {\bibfnamefont {Y.-L.}\ \bibnamefont {Xu}},
  \bibinfo {author} {\bibfnamefont {W.~S.}\ \bibnamefont {Fegadolli}}, \bibinfo
  {author} {\bibfnamefont {M.-H.}\ \bibnamefont {Lu}}, \bibinfo {author}
  {\bibfnamefont {J.~E.~B.}\ \bibnamefont {Oliveira}}, \bibinfo {author}
  {\bibfnamefont {V.~R.}\ \bibnamefont {Almeida}}, \bibinfo {author}
  {\bibfnamefont {Y.-F.}\ \bibnamefont {Chen}},\ and\ \bibinfo {author}
  {\bibfnamefont {A.}~\bibnamefont {Scherer}},\ }\bibfield  {title} {\bibinfo
  {title} {Experimental demonstration of a unidirectional reflectionless
  parity-time metamaterial at optical frequencies},\ }\href
  {https://doi.org/10.1038/nmat3495} {\bibfield  {journal} {\bibinfo  {journal}
  {Nature Materials}\ }\textbf {\bibinfo {volume} {12}},\ \bibinfo {pages}
  {108} (\bibinfo {year} {2013})}\BibitemShut {NoStop}%
\bibitem [{\citenamefont {Feng}\ \emph {et~al.}(2014)\citenamefont {Feng},
  \citenamefont {Wong}, \citenamefont {Ma}, \citenamefont {Wang},\ and\
  \citenamefont {Zhang}}]{feng2014}%
  \BibitemOpen
  \bibfield  {author} {\bibinfo {author} {\bibfnamefont {L.}~\bibnamefont
  {Feng}}, \bibinfo {author} {\bibfnamefont {Z.~J.}\ \bibnamefont {Wong}},
  \bibinfo {author} {\bibfnamefont {R.-M.}\ \bibnamefont {Ma}}, \bibinfo
  {author} {\bibfnamefont {Y.}~\bibnamefont {Wang}},\ and\ \bibinfo {author}
  {\bibfnamefont {X.}~\bibnamefont {Zhang}},\ }\bibfield  {title} {\bibinfo
  {title} {Single-mode laser by parity-time symmetry breaking},\ }\href
  {https://doi.org/10.1126/science.1258479} {\bibfield  {journal} {\bibinfo
  {journal} {Science}\ }\textbf {\bibinfo {volume} {346}},\ \bibinfo {pages}
  {972} (\bibinfo {year} {2014})}\BibitemShut {NoStop}%
\bibitem [{\citenamefont {Hodaei}\ \emph {et~al.}(2014)\citenamefont {Hodaei},
  \citenamefont {Miri}, \citenamefont {Heinrich}, \citenamefont
  {Christodoulides},\ and\ \citenamefont {Khajavikhan}}]{hodaei2014}%
  \BibitemOpen
  \bibfield  {author} {\bibinfo {author} {\bibfnamefont {H.}~\bibnamefont
  {Hodaei}}, \bibinfo {author} {\bibfnamefont {M.-A.}\ \bibnamefont {Miri}},
  \bibinfo {author} {\bibfnamefont {M.}~\bibnamefont {Heinrich}}, \bibinfo
  {author} {\bibfnamefont {D.~N.}\ \bibnamefont {Christodoulides}},\ and\
  \bibinfo {author} {\bibfnamefont {M.}~\bibnamefont {Khajavikhan}},\
  }\bibfield  {title} {\bibinfo {title} {Parity-time-symmetric microring
  lasers},\ }\href {https://doi.org/10.1126/science.1258480} {\bibfield
  {journal} {\bibinfo  {journal} {Science}\ }\textbf {\bibinfo {volume}
  {346}},\ \bibinfo {pages} {975} (\bibinfo {year} {2014})}\BibitemShut
  {NoStop}%
\bibitem [{\citenamefont {Peng}\ \emph {et~al.}(2014)\citenamefont {Peng},
  \citenamefont {{\"{O}}zdemir}, \citenamefont {Rotter}, \citenamefont
  {Yilmaz}, \citenamefont {Liertzer}, \citenamefont {Monifi}, \citenamefont
  {Bender}, \citenamefont {Nori},\ and\ \citenamefont {Yang}}]{peng2014}%
  \BibitemOpen
  \bibfield  {author} {\bibinfo {author} {\bibfnamefont {B.}~\bibnamefont
  {Peng}}, \bibinfo {author} {\bibfnamefont {{\c {S}}.~K.}\ \bibnamefont
  {{\"{O}}zdemir}}, \bibinfo {author} {\bibfnamefont {S.}~\bibnamefont
  {Rotter}}, \bibinfo {author} {\bibfnamefont {H.}~\bibnamefont {Yilmaz}},
  \bibinfo {author} {\bibfnamefont {M.}~\bibnamefont {Liertzer}}, \bibinfo
  {author} {\bibfnamefont {F.}~\bibnamefont {Monifi}}, \bibinfo {author}
  {\bibfnamefont {C.~M.}\ \bibnamefont {Bender}}, \bibinfo {author}
  {\bibfnamefont {F.}~\bibnamefont {Nori}},\ and\ \bibinfo {author}
  {\bibfnamefont {L.}~\bibnamefont {Yang}},\ }\bibfield  {title} {\bibinfo
  {title} {Loss-induced suppression and revival of lasing},\ }\href
  {https://doi.org/10.1126/science.1258004} {\bibfield  {journal} {\bibinfo
  {journal} {Science}\ }\textbf {\bibinfo {volume} {346}},\ \bibinfo {pages}
  {328} (\bibinfo {year} {2014})}\BibitemShut {NoStop}%
\bibitem [{\citenamefont {Weimann}\ \emph {et~al.}(2017)\citenamefont
  {Weimann}, \citenamefont {Kremer}, \citenamefont {Plotnik}, \citenamefont
  {Lumer}, \citenamefont {Nolte}, \citenamefont {Makris}, \citenamefont
  {Segev}, \citenamefont {Rechtsman},\ and\ \citenamefont
  {Szameit}}]{weimann2017}%
  \BibitemOpen
  \bibfield  {author} {\bibinfo {author} {\bibfnamefont {S.}~\bibnamefont
  {Weimann}}, \bibinfo {author} {\bibfnamefont {M.}~\bibnamefont {Kremer}},
  \bibinfo {author} {\bibfnamefont {Y.}~\bibnamefont {Plotnik}}, \bibinfo
  {author} {\bibfnamefont {Y.}~\bibnamefont {Lumer}}, \bibinfo {author}
  {\bibfnamefont {S.}~\bibnamefont {Nolte}}, \bibinfo {author} {\bibfnamefont
  {K.~G.}\ \bibnamefont {Makris}}, \bibinfo {author} {\bibfnamefont
  {M.}~\bibnamefont {Segev}}, \bibinfo {author} {\bibfnamefont {M.~{\^A}.~C.}\
  \bibnamefont {Rechtsman}},\ and\ \bibinfo {author} {\bibfnamefont
  {A.}~\bibnamefont {Szameit}},\ }\bibfield  {title} {\bibinfo {title}
  {Topologically protected bound states in photonic parity-time-symmetric
  crystals},\ }\href {https://doi.org/10.1038/nmat4811} {\bibfield  {journal}
  {\bibinfo  {journal} {Nature Materials}\ }\textbf {\bibinfo {volume} {16}},\
  \bibinfo {pages} {433} (\bibinfo {year} {2017})}\BibitemShut {NoStop}%
\bibitem [{\citenamefont {Miri}\ and\ \citenamefont {Alù}(2019)}]{miri2019}%
  \BibitemOpen
  \bibfield  {author} {\bibinfo {author} {\bibfnamefont {M.-A.}\ \bibnamefont
  {Miri}}\ and\ \bibinfo {author} {\bibfnamefont {A.}~\bibnamefont {Alù}},\
  }\bibfield  {title} {\bibinfo {title} {Exceptional points in optics and
  photonics},\ }\href {https://doi.org/10.1126/science.aar7709} {\bibfield
  {journal} {\bibinfo  {journal} {Science}\ }\textbf {\bibinfo {volume}
  {363}},\ \bibinfo {pages} {eaar7709} (\bibinfo {year} {2019})}\BibitemShut
  {NoStop}%
\bibitem [{\citenamefont {{\"{O}}zdemir}\ \emph {et~al.}(2019)\citenamefont
  {{\"{O}}zdemir}, \citenamefont {Rotter}, \citenamefont {Nori},\ and\
  \citenamefont {Yang}}]{ozdemir2019}%
  \BibitemOpen
  \bibfield  {author} {\bibinfo {author} {\bibfnamefont {{\c {S}}.~K.}\
  \bibnamefont {{\"{O}}zdemir}}, \bibinfo {author} {\bibfnamefont
  {S.}~\bibnamefont {Rotter}}, \bibinfo {author} {\bibfnamefont
  {F.}~\bibnamefont {Nori}},\ and\ \bibinfo {author} {\bibfnamefont
  {L.}~\bibnamefont {Yang}},\ }\bibfield  {title} {\bibinfo {title}
  {Parity-time symmetry and exceptional points in photonics},\ }\href
  {https://doi.org/10.1038/s41563-019-0304-9} {\bibfield  {journal} {\bibinfo
  {journal} {Nature Materials}\ }\textbf {\bibinfo {volume} {18}},\ \bibinfo
  {pages} {783} (\bibinfo {year} {2019})}\BibitemShut {NoStop}%
\bibitem [{\citenamefont {Xiao}\ \emph {et~al.}(2020)\citenamefont {Xiao},
  \citenamefont {Deng}, \citenamefont {Wang}, \citenamefont {Zhu},
  \citenamefont {Wang}, \citenamefont {Yi},\ and\ \citenamefont
  {Xue}}]{xiao2020}%
  \BibitemOpen
  \bibfield  {author} {\bibinfo {author} {\bibfnamefont {L.}~\bibnamefont
  {Xiao}}, \bibinfo {author} {\bibfnamefont {T.}~\bibnamefont {Deng}}, \bibinfo
  {author} {\bibfnamefont {K.}~\bibnamefont {Wang}}, \bibinfo {author}
  {\bibfnamefont {G.}~\bibnamefont {Zhu}}, \bibinfo {author} {\bibfnamefont
  {Z.}~\bibnamefont {Wang}}, \bibinfo {author} {\bibfnamefont {W.}~\bibnamefont
  {Yi}},\ and\ \bibinfo {author} {\bibfnamefont {P.}~\bibnamefont {Xue}},\
  }\bibfield  {title} {\bibinfo {title} {Non-hermitian bulk-boundary
  correspondence in quantum dynamics},\ }\href
  {https://doi.org/10.1038/s41567-020-0836-6} {\bibfield  {journal} {\bibinfo
  {journal} {Nature Physics}\ }\textbf {\bibinfo {volume} {16}},\ \bibinfo
  {pages} {761} (\bibinfo {year} {2020})}\BibitemShut {NoStop}%
\bibitem [{\citenamefont {Bergholtz}\ \emph {et~al.}(2021)\citenamefont
  {Bergholtz}, \citenamefont {Budich},\ and\ \citenamefont
  {Kunst}}]{bergholtz2021}%
  \BibitemOpen
  \bibfield  {author} {\bibinfo {author} {\bibfnamefont {E.~J.}\ \bibnamefont
  {Bergholtz}}, \bibinfo {author} {\bibfnamefont {J.~C.}\ \bibnamefont
  {Budich}},\ and\ \bibinfo {author} {\bibfnamefont {F.~K.}\ \bibnamefont
  {Kunst}},\ }\bibfield  {title} {\bibinfo {title} {Exceptional topology of
  non-hermitian systems},\ }\href
  {https://doi.org/10.1103/RevModPhys.93.015005} {\bibfield  {journal}
  {\bibinfo  {journal} {Rev. Mod. Phys.}\ }\textbf {\bibinfo {volume} {93}},\
  \bibinfo {pages} {015005} (\bibinfo {year} {2021})}\BibitemShut {NoStop}%
\bibitem [{\citenamefont {Heiss}(1999)}]{heiss1999}%
  \BibitemOpen
  \bibfield  {author} {\bibinfo {author} {\bibfnamefont {W.~D.}\ \bibnamefont
  {Heiss}},\ }\bibfield  {title} {\bibinfo {title} {Phases of wave functions
  and level repulsion},\ }\href {https://doi.org/10.1007/s100530050339}
  {\bibfield  {journal} {\bibinfo  {journal} {The European Physical Journal D -
  Atomic, Molecular, Optical and Plasma Physics}\ }\textbf {\bibinfo {volume}
  {7}},\ \bibinfo {pages} {1} (\bibinfo {year} {1999})}\BibitemShut {NoStop}%
\bibitem [{\citenamefont {Heiss}(2000)}]{heiss2000}%
  \BibitemOpen
  \bibfield  {author} {\bibinfo {author} {\bibfnamefont {W.~D.}\ \bibnamefont
  {Heiss}},\ }\bibfield  {title} {\bibinfo {title} {Repulsion of resonance
  states and exceptional points},\ }\href
  {https://doi.org/10.1103/PhysRevE.61.929} {\bibfield  {journal} {\bibinfo
  {journal} {Phys. Rev. E}\ }\textbf {\bibinfo {volume} {61}},\ \bibinfo
  {pages} {929} (\bibinfo {year} {2000})}\BibitemShut {NoStop}%
\bibitem [{\citenamefont {Keck}\ \emph {et~al.}(2003)\citenamefont {Keck},
  \citenamefont {Korsch},\ and\ \citenamefont {Mossmann}}]{keck2003}%
  \BibitemOpen
  \bibfield  {author} {\bibinfo {author} {\bibfnamefont {F.}~\bibnamefont
  {Keck}}, \bibinfo {author} {\bibfnamefont {H.~J.}\ \bibnamefont {Korsch}},\
  and\ \bibinfo {author} {\bibfnamefont {S.}~\bibnamefont {Mossmann}},\
  }\bibfield  {title} {\bibinfo {title} {Unfolding a diabolic point: a
  generalized crossing scenario},\ }\href
  {https://doi.org/10.1088/0305-4470/36/8/310} {\bibfield  {journal} {\bibinfo
  {journal} {Journal of Physics A: Mathematical and General}\ }\textbf
  {\bibinfo {volume} {36}},\ \bibinfo {pages} {2125} (\bibinfo {year}
  {2003})}\BibitemShut {NoStop}%
\bibitem [{\citenamefont {Berry}\ and\ \citenamefont
  {Uzdin}(2011)}]{berry2011}%
  \BibitemOpen
  \bibfield  {author} {\bibinfo {author} {\bibfnamefont {M.~V.}\ \bibnamefont
  {Berry}}\ and\ \bibinfo {author} {\bibfnamefont {R.}~\bibnamefont {Uzdin}},\
  }\bibfield  {title} {\bibinfo {title} {Slow non-hermitian cycling: exact
  solutions and the stokes phenomenon},\ }\href
  {https://doi.org/10.1088/1751-8113/44/43/435303} {\bibfield  {journal}
  {\bibinfo  {journal} {Journal of Physics A: Mathematical and Theoretical}\
  }\textbf {\bibinfo {volume} {44}},\ \bibinfo {pages} {435303} (\bibinfo
  {year} {2011})}\BibitemShut {NoStop}%
\bibitem [{\citenamefont {Uzdin}\ \emph {et~al.}(2011)\citenamefont {Uzdin},
  \citenamefont {Mailybaev},\ and\ \citenamefont {Moiseyev}}]{uzdin2011}%
  \BibitemOpen
  \bibfield  {author} {\bibinfo {author} {\bibfnamefont {R.}~\bibnamefont
  {Uzdin}}, \bibinfo {author} {\bibfnamefont {A.}~\bibnamefont {Mailybaev}},\
  and\ \bibinfo {author} {\bibfnamefont {N.}~\bibnamefont {Moiseyev}},\
  }\bibfield  {title} {\bibinfo {title} {On the observability and asymmetry of
  adiabatic state flips generated by exceptional points},\ }\href
  {https://doi.org/10.1088/1751-8113/44/43/435302} {\bibfield  {journal}
  {\bibinfo  {journal} {Journal of Physics A: Mathematical and Theoretical}\
  }\textbf {\bibinfo {volume} {44}},\ \bibinfo {pages} {435302} (\bibinfo
  {year} {2011})}\BibitemShut {NoStop}%
\bibitem [{\citenamefont {Milburn}\ \emph {et~al.}(2015)\citenamefont
  {Milburn}, \citenamefont {Doppler}, \citenamefont {Holmes}, \citenamefont
  {Portolan}, \citenamefont {Rotter},\ and\ \citenamefont
  {Rabl}}]{milburn2015}%
  \BibitemOpen
  \bibfield  {author} {\bibinfo {author} {\bibfnamefont {T.~J.}\ \bibnamefont
  {Milburn}}, \bibinfo {author} {\bibfnamefont {J.}~\bibnamefont {Doppler}},
  \bibinfo {author} {\bibfnamefont {C.~A.}\ \bibnamefont {Holmes}}, \bibinfo
  {author} {\bibfnamefont {S.}~\bibnamefont {Portolan}}, \bibinfo {author}
  {\bibfnamefont {S.}~\bibnamefont {Rotter}},\ and\ \bibinfo {author}
  {\bibfnamefont {P.}~\bibnamefont {Rabl}},\ }\bibfield  {title} {\bibinfo
  {title} {General description of quasiadiabatic dynamical phenomena near
  exceptional points},\ }\href {https://doi.org/10.1103/PhysRevA.92.052124}
  {\bibfield  {journal} {\bibinfo  {journal} {Phys. Rev. A}\ }\textbf {\bibinfo
  {volume} {92}},\ \bibinfo {pages} {052124} (\bibinfo {year}
  {2015})}\BibitemShut {NoStop}%
\bibitem [{\citenamefont {Dembowski}\ \emph {et~al.}(2001)\citenamefont
  {Dembowski}, \citenamefont {Gr{\"{a}}f}, \citenamefont {Harney},
  \citenamefont {Heine}, \citenamefont {Heiss}, \citenamefont {Rehfeld},\ and\
  \citenamefont {Richter}}]{dembowski2001}%
  \BibitemOpen
  \bibfield  {author} {\bibinfo {author} {\bibfnamefont {C.}~\bibnamefont
  {Dembowski}}, \bibinfo {author} {\bibfnamefont {H.-D.}\ \bibnamefont
  {Gr{\"{a}}f}}, \bibinfo {author} {\bibfnamefont {H.~L.}\ \bibnamefont
  {Harney}}, \bibinfo {author} {\bibfnamefont {A.}~\bibnamefont {Heine}},
  \bibinfo {author} {\bibfnamefont {W.~D.}\ \bibnamefont {Heiss}}, \bibinfo
  {author} {\bibfnamefont {H.}~\bibnamefont {Rehfeld}},\ and\ \bibinfo {author}
  {\bibfnamefont {A.}~\bibnamefont {Richter}},\ }\bibfield  {title} {\bibinfo
  {title} {Experimental observation of the topological structure of exceptional
  points},\ }\href {https://doi.org/10.1103/PhysRevLett.86.787} {\bibfield
  {journal} {\bibinfo  {journal} {Phys. Rev. Lett.}\ }\textbf {\bibinfo
  {volume} {86}},\ \bibinfo {pages} {787} (\bibinfo {year} {2001})}\BibitemShut
  {NoStop}%
\bibitem [{\citenamefont {Xu}\ \emph {et~al.}(2016)\citenamefont {Xu},
  \citenamefont {Mason}, \citenamefont {Jiang},\ and\ \citenamefont
  {Harris}}]{xu2016}%
  \BibitemOpen
  \bibfield  {author} {\bibinfo {author} {\bibfnamefont {H.}~\bibnamefont
  {Xu}}, \bibinfo {author} {\bibfnamefont {D.}~\bibnamefont {Mason}}, \bibinfo
  {author} {\bibfnamefont {L.}~\bibnamefont {Jiang}},\ and\ \bibinfo {author}
  {\bibfnamefont {J.~G.~E.}\ \bibnamefont {Harris}},\ }\bibfield  {title}
  {\bibinfo {title} {Topological energy transfer in an optomechanical system
  with exceptional points},\ }\href {https://doi.org/10.1038/nature18604}
  {\bibfield  {journal} {\bibinfo  {journal} {Nature}\ }\textbf {\bibinfo
  {volume} {537}},\ \bibinfo {pages} {80} (\bibinfo {year} {2016})}\BibitemShut
  {NoStop}%
\bibitem [{\citenamefont {Doppler}\ \emph {et~al.}(2016)\citenamefont
  {Doppler}, \citenamefont {Mailybaev}, \citenamefont {B{\"o}hm}, \citenamefont
  {Kuhl}, \citenamefont {Girschik}, \citenamefont {Libisch}, \citenamefont
  {Milburn}, \citenamefont {Rabl}, \citenamefont {Moiseyev},\ and\
  \citenamefont {Rotter}}]{doppler2016}%
  \BibitemOpen
  \bibfield  {author} {\bibinfo {author} {\bibfnamefont {J.}~\bibnamefont
  {Doppler}}, \bibinfo {author} {\bibfnamefont {A.~A.}\ \bibnamefont
  {Mailybaev}}, \bibinfo {author} {\bibfnamefont {J.}~\bibnamefont {B{\"o}hm}},
  \bibinfo {author} {\bibfnamefont {U.}~\bibnamefont {Kuhl}}, \bibinfo {author}
  {\bibfnamefont {A.}~\bibnamefont {Girschik}}, \bibinfo {author}
  {\bibfnamefont {F.}~\bibnamefont {Libisch}}, \bibinfo {author} {\bibfnamefont
  {T.~J.}\ \bibnamefont {Milburn}}, \bibinfo {author} {\bibfnamefont
  {P.}~\bibnamefont {Rabl}}, \bibinfo {author} {\bibfnamefont {N.}~\bibnamefont
  {Moiseyev}},\ and\ \bibinfo {author} {\bibfnamefont {S.}~\bibnamefont
  {Rotter}},\ }\bibfield  {title} {\bibinfo {title} {Dynamically encircling an
  exceptional point for asymmetric mode switching},\ }\href
  {https://doi.org/10.1038/nature18605} {\bibfield  {journal} {\bibinfo
  {journal} {Nature}\ }\textbf {\bibinfo {volume} {537}},\ \bibinfo {pages}
  {76} (\bibinfo {year} {2016})}\BibitemShut {NoStop}%
\bibitem [{\citenamefont {Liu}\ \emph {et~al.}(2021)\citenamefont {Liu},
  \citenamefont {Wu}, \citenamefont {Duan}, \citenamefont {Rong},\ and\
  \citenamefont {Du}}]{liu2021}%
  \BibitemOpen
  \bibfield  {author} {\bibinfo {author} {\bibfnamefont {W.}~\bibnamefont
  {Liu}}, \bibinfo {author} {\bibfnamefont {Y.}~\bibnamefont {Wu}}, \bibinfo
  {author} {\bibfnamefont {C.-K.}\ \bibnamefont {Duan}}, \bibinfo {author}
  {\bibfnamefont {X.}~\bibnamefont {Rong}},\ and\ \bibinfo {author}
  {\bibfnamefont {J.}~\bibnamefont {Du}},\ }\bibfield  {title} {\bibinfo
  {title} {Dynamically encircling an exceptional point in a real quantum
  system},\ }\href {https://doi.org/10.1103/PhysRevLett.126.170506} {\bibfield
  {journal} {\bibinfo  {journal} {Phys. Rev. Lett.}\ }\textbf {\bibinfo
  {volume} {126}},\ \bibinfo {pages} {170506} (\bibinfo {year}
  {2021})}\BibitemShut {NoStop}%
\bibitem [{\citenamefont {Li}\ \emph {et~al.}(2009)\citenamefont {Li},
  \citenamefont {Chu}, \citenamefont {Jain},\ and\ \citenamefont
  {Shen}}]{li2009}%
  \BibitemOpen
  \bibfield  {author} {\bibinfo {author} {\bibfnamefont {J.}~\bibnamefont
  {Li}}, \bibinfo {author} {\bibfnamefont {R.-L.}\ \bibnamefont {Chu}},
  \bibinfo {author} {\bibfnamefont {J.~K.}\ \bibnamefont {Jain}},\ and\
  \bibinfo {author} {\bibfnamefont {S.-Q.}\ \bibnamefont {Shen}},\ }\bibfield
  {title} {\bibinfo {title} {Topological anderson insulator},\ }\href
  {https://doi.org/10.1103/PhysRevLett.102.136806} {\bibfield  {journal}
  {\bibinfo  {journal} {Phys. Rev. Lett.}\ }\textbf {\bibinfo {volume} {102}},\
  \bibinfo {pages} {136806} (\bibinfo {year} {2009})}\BibitemShut {NoStop}%
\bibitem [{\citenamefont {P\'erez-Gonz\'alez}\ \emph
  {et~al.}(2019)\citenamefont {P\'erez-Gonz\'alez}, \citenamefont {Bello},
  \citenamefont {G\'omez-Le\'on},\ and\ \citenamefont
  {Platero}}]{gonzalez2019}%
  \BibitemOpen
  \bibfield  {author} {\bibinfo {author} {\bibfnamefont {B.}~\bibnamefont
  {P\'erez-Gonz\'alez}}, \bibinfo {author} {\bibfnamefont {M.}~\bibnamefont
  {Bello}}, \bibinfo {author} {\bibfnamefont {A.}~\bibnamefont
  {G\'omez-Le\'on}},\ and\ \bibinfo {author} {\bibfnamefont {G.}~\bibnamefont
  {Platero}},\ }\bibfield  {title} {\bibinfo {title} {Interplay between
  long-range hopping and disorder in topological systems},\ }\href
  {https://doi.org/10.1103/PhysRevB.99.035146} {\bibfield  {journal} {\bibinfo
  {journal} {Phys. Rev. B}\ }\textbf {\bibinfo {volume} {99}},\ \bibinfo
  {pages} {035146} (\bibinfo {year} {2019})}\BibitemShut {NoStop}%
\bibitem [{\citenamefont {Ribeiro}\ \emph {et~al.}(2017)\citenamefont
  {Ribeiro}, \citenamefont {Baksic},\ and\ \citenamefont
  {Clerk}}]{ribeiro2017}%
  \BibitemOpen
  \bibfield  {author} {\bibinfo {author} {\bibfnamefont {H.}~\bibnamefont
  {Ribeiro}}, \bibinfo {author} {\bibfnamefont {A.}~\bibnamefont {Baksic}},\
  and\ \bibinfo {author} {\bibfnamefont {A.~A.}\ \bibnamefont {Clerk}},\
  }\bibfield  {title} {\bibinfo {title} {Systematic magnus-based approach for
  suppressing leakage and nonadiabatic errors in quantum dynamics},\ }\href
  {https://doi.org/10.1103/PhysRevX.7.011021} {\bibfield  {journal} {\bibinfo
  {journal} {Phys. Rev. X}\ }\textbf {\bibinfo {volume} {7}},\ \bibinfo {pages}
  {011021} (\bibinfo {year} {2017})}\BibitemShut {NoStop}%
\bibitem [{\citenamefont {Figueiredo~Roque}\ \emph {et~al.}(2021)\citenamefont
  {Figueiredo~Roque}, \citenamefont {Clerk},\ and\ \citenamefont
  {Ribeiro}}]{roque2021}%
  \BibitemOpen
  \bibfield  {author} {\bibinfo {author} {\bibfnamefont {T.}~\bibnamefont
  {Figueiredo~Roque}}, \bibinfo {author} {\bibfnamefont {A.~A.}\ \bibnamefont
  {Clerk}},\ and\ \bibinfo {author} {\bibfnamefont {H.}~\bibnamefont
  {Ribeiro}},\ }\bibfield  {title} {\bibinfo {title} {Engineering fast
  high-fidelity quantum operations with constrained interactions},\ }\href
  {https://doi.org/10.1038/s41534-020-00349-z} {\bibfield  {journal} {\bibinfo
  {journal} {npj Quantum Information}\ }\textbf {\bibinfo {volume} {7}},\
  \bibinfo {pages} {28} (\bibinfo {year} {2021})}\BibitemShut {NoStop}%
\bibitem [{\citenamefont {Magnus}(1954)}]{magnus1954}%
  \BibitemOpen
  \bibfield  {author} {\bibinfo {author} {\bibfnamefont {W.}~\bibnamefont
  {Magnus}},\ }\bibfield  {title} {\bibinfo {title} {On the exponential
  solution of differential equations for a linear operator},\ }\href
  {https://doi.org/10.1002/cpa.3160070404} {\bibfield  {journal} {\bibinfo
  {journal} {Communications on Pure and Applied Mathematics}\ }\textbf
  {\bibinfo {volume} {7}},\ \bibinfo {pages} {649} (\bibinfo {year}
  {1954})}\BibitemShut {NoStop}%
\bibitem [{\citenamefont {Blanes}\ \emph {et~al.}(2009)\citenamefont {Blanes},
  \citenamefont {Casas}, \citenamefont {Oteo},\ and\ \citenamefont
  {Ros}}]{blanes2009}%
  \BibitemOpen
  \bibfield  {author} {\bibinfo {author} {\bibfnamefont {S.}~\bibnamefont
  {Blanes}}, \bibinfo {author} {\bibfnamefont {F.}~\bibnamefont {Casas}},
  \bibinfo {author} {\bibfnamefont {J.~A.}\ \bibnamefont {Oteo}},\ and\
  \bibinfo {author} {\bibfnamefont {J.}~\bibnamefont {Ros}},\ }\bibfield
  {title} {\bibinfo {title} {The magnus expansion and some of its
  applications},\ }\href
  {https://doi.org/http://dx.doi.org/10.1016/j.physrep.2008.11.001} {\bibfield
  {journal} {\bibinfo  {journal} {Physics Reports}\ }\textbf {\bibinfo {volume}
  {470}},\ \bibinfo {pages} {151 } (\bibinfo {year} {2009})}\BibitemShut
  {NoStop}%
\bibitem [{\citenamefont {Born}\ and\ \citenamefont {Fock}(1928)}]{born1928}%
  \BibitemOpen
  \bibfield  {author} {\bibinfo {author} {\bibfnamefont {M.}~\bibnamefont
  {Born}}\ and\ \bibinfo {author} {\bibfnamefont {V.}~\bibnamefont {Fock}},\
  }\bibfield  {title} {\bibinfo {title} {Beweis des adiabatensatzes},\ }\href
  {https://doi.org/10.1007/BF01343193} {\bibfield  {journal} {\bibinfo
  {journal} {Zeitschrift f{\"{u}}r Physik}\ }\textbf {\bibinfo {volume} {51}},\
  \bibinfo {pages} {165} (\bibinfo {year} {1928})}\BibitemShut {NoStop}%
\bibitem [{\citenamefont {Dyson}(1949)}]{dyson1949}%
  \BibitemOpen
  \bibfield  {author} {\bibinfo {author} {\bibfnamefont {F.~J.}\ \bibnamefont
  {Dyson}},\ }\bibfield  {title} {\bibinfo {title} {The radiation theories of
  tomonaga, schwinger, and feynman},\ }\href
  {https://doi.org/10.1103/PhysRev.75.486} {\bibfield  {journal} {\bibinfo
  {journal} {Phys. Rev.}\ }\textbf {\bibinfo {volume} {75}},\ \bibinfo {pages}
  {486} (\bibinfo {year} {1949})}\BibitemShut {NoStop}%
\bibitem [{\citenamefont {Wiebe}\ and\ \citenamefont
  {Babcock}(2012)}]{wiebe2012}%
  \BibitemOpen
  \bibfield  {author} {\bibinfo {author} {\bibfnamefont {N.}~\bibnamefont
  {Wiebe}}\ and\ \bibinfo {author} {\bibfnamefont {N.~S.}\ \bibnamefont
  {Babcock}},\ }\bibfield  {title} {\bibinfo {title} {Improved error-scaling
  for adiabatic quantum evolutions},\ }\href
  {https://doi.org/10.1088/1367-2630/14/1/013024} {\bibfield  {journal}
  {\bibinfo  {journal} {New Journal of Physics}\ }\textbf {\bibinfo {volume}
  {14}},\ \bibinfo {pages} {013024} (\bibinfo {year} {2012})}\BibitemShut
  {NoStop}%
\bibitem [{\citenamefont {Ribeiro}\ and\ \citenamefont
  {Clerk}(2019)}]{ribeiro2019}%
  \BibitemOpen
  \bibfield  {author} {\bibinfo {author} {\bibfnamefont {H.}~\bibnamefont
  {Ribeiro}}\ and\ \bibinfo {author} {\bibfnamefont {A.~A.}\ \bibnamefont
  {Clerk}},\ }\bibfield  {title} {\bibinfo {title} {Accelerated adiabatic
  quantum gates: Optimizing speed versus robustness},\ }\href
  {https://doi.org/10.1103/PhysRevA.100.032323} {\bibfield  {journal} {\bibinfo
   {journal} {Phys. Rev. A}\ }\textbf {\bibinfo {volume} {100}},\ \bibinfo
  {pages} {032323} (\bibinfo {year} {2019})}\BibitemShut {NoStop}%
\bibitem [{Note1()}]{Note1}%
  \BibitemOpen
  \bibinfo {note} {For $n=0$, we define $D_\protect \mathrm {mod,I}^{(0)} (t) =
  V_\protect \mathrm {good,I} (t) + V_\protect \mathrm {bad,I}
  (t)$.}\BibitemShut {Stop}%
\end{thebibliography}
\end{document}